\documentclass[reprint,aps,prd,nofootinbib]{revtex4-2}
\usepackage{graphicx}
\usepackage{dcolumn}
\usepackage{bm,amsmath,amssymb}
\newcommand{\pd}[3]{\frac{\partial^{#3} #1}{\partial {#2}^{#3}}} 
\newcommand{\tr}{\tilde{r}}
\newcommand{\tE}{\tilde{E}}

\usepackage{orcidlink}
\usepackage{natbib}

\begin{document}

\preprint{APS/123-QED}

\title{A radio-frequency WIMP search with the MeerKAT Galaxy Cluster Legacy Survey} 

\author{Natasha Lavis \orcidlink{https://orcid.org/	
0000-0002-4116-5480}}%
\author{Michael Sarkis \orcidlink{https://orcid.org/0000-0002-7935-8608}}
\author{Geoff Beck \orcidlink{https://orcid.org/0000-0003-4916-4914} }
\altaffiliation{geoffrey.beck@wits.ac.za}
 
\affiliation{School of Physics and Centre for Astrophysics, University of the Witwatersrand, Johannesburg, Wits 2050, South Africa
}%
\author{Kenda Knowles \orcidlink{https://orcid.org/0000-0002-8452-0825}}
\affiliation{Centre for Radio Astronomy Techniques and Technologies, Department of Physics and Electronics, Rhodes University, P.O. Box 94,
Makhanda 6140, South Africa}

\date{\today}

\begin{abstract}
Radio-frequency, indirect dark matter searches have recently been gaining prevalence, due to the high sensitivity and resolution capabilities of the new generation of radio interferometers. MeerKAT is currently one of the most sensitive instruments of its kind, making it ideal for indirect dark matter searches. By making use of publicly available data from the MeerKAT Galaxy Cluster Legacy Survey we are able to use both the observed diffuse synchrotron emission and non-detections to constrain the WIMP dark matter parameter space. In addition to a subset of generic WIMP annihilation channels, we probe the dark matter candidate within the 2HDM+S particle physics model, which was developed as an explanation for anomalies observed in the Large Hadron Collider data from runs 1 and 2. By undertaking a statistical analysis of the radio flux densities within galaxy clusters we are able to exclude the thermal relic value for WIMP masses $< \, \sim 1000 $ GeV for annihilation into bottom quarks with our median case. This is competitive with the best constraints in the current literature.
\end{abstract}

\maketitle

\section{Introduction}
The unknown nature of dark matter (DM) remains an eyesore in our current cosmological paradigm. The nature of DM can be probed through indirect searches, which look for visible products of the annihilation or decay of a DM candidate. Such searches are particularly promising for Weakly Interacting Massive Particles (WIMPs), the most investigated class of DM candidates~\cite{roszkowski2018wimp}. Historically, these searches have been dominated by gamma-ray experiments, such as Fermi-LAT~\cite{albert2017fermi,Hoof_2020}, due to their low attenuation, high detection efficiency, and simplicity of the predicted signal. For WIMP models with masses above a few GeV, the indirect emission would be in the form of long-lived leptons which then produce synchrotron radiation when interacting with the object's magnetic field ~\cite{beck2019radio}. In the case that the leptons, or rather electrons, produced in an annihilation have energies roughly below 10 GeV, the resulting synchrotron emission will be in the radio band. Indirect searches in the radio band were previously disfavored, due to complicating factors such as the strong dependence on the magnetic field configuration within the target object and the diffusion of the charged particles, these being generally not well known in most astrophysical structures.  With the new generation of radio interferometers, such as MeerKAT~\cite{jonas2016meerkat} a precursor to the Square Kilometer Array (SKA)~\cite{dewdney2009square}, radio band DM searches are gaining in prevalence. This is due to their superior angular resolution to gamma-ray telescopes, imperative to limit confusion between diffuse DM emission and unresolved point sources, as well as a high sensitivity to faint flux levels expected for diffuse DM emission. The advancements in radio astronomy techniques and technology are beginning to overcome the traditional obstacles in radio DM searches. This is typified by a recent result that demonstrates that ASKAP observations of the Large Magellanic Cloud are capable of ruling out WIMPs annihilating in $b$-quarks for masses $\lesssim 700$ GeV~\cite{regis2021emu}, far in excess of Fermi-LAT gamma-ray results in dwarf spheroidal galaxies~\cite{Hoof_2020}. 

MeerKAT is the precursor for the Square Kilometer Array (SKA) mid-frequency array and is currently the most sensitive decimetre wavelength radio interferometer array in the world. Operating with 64 antennas of 13.5 m diameter with an offset-Gregorian feed. The array configuration allows for exceptional simultaneous sensitivity to a wide range of angular scales \cite{article}. 

DM searches performed with MeerKAT will benefit from the higher sensitivity available than prior searches performed with instruments such as Green Bank Telescope (GBT) and the Australia Telescope Compact Array (ATCA) \cite{natarajan2015green, regis2017dark, kar2019constraints}. The improved sensitivity is able to observe fainter diffuse sources, expanding the DM parameter space that can be probed. The high angular resolution provided by the long baselines allows for the simultaneous detection of large and small-scale sources, thus reducing confusion caused by unresolved point sources. 

Previous radio frequency DM searches have focused on nearby dwarf spheroidal galaxies (dSph), for example: \cite{colafrancesco2006multi,Siffert_2010,spekkens2013deep,regis2017dark,kar2019constraints,Vollmann_2021,basu2021stringent,regis2021emu,Gajovi__2023}. Dwarf galaxies are DM-dominated and benefit from low baryonic background emissions. However, the magnetic field and diffusive uncertainties are large and strongly affect the predicted surface brightnesses.  Investigations of larger structures, on the scale of galaxy clusters, have the advantage that the effect of diffusion is less significant. Clusters of galaxies are promising targets for radio frequency DM searches, as they are massive, DM-dominated, and are known to host $\mu$G-scale magnetic fields. In addition to this, clusters are far less sensitive to uncertainties in the predicted signal induced by the diffusion processes, as a consequence of their large scale. The potential of galaxy clusters as target objects has been demonstrated with forecasting of limits with high-resolution observations of well-known clusters~\cite{beck2022galaxy}.   

In this work, we compare predicted DM fluxes within galaxy clusters to the diffuse flux measured with MeerKAT to produce limits for the annihilation cross-section across a range of DM masses, with various methods of analysis. This is achieved using MeerKAT Galaxy Cluster Legacy Survey~\cite{knowles2022meerkat} (MGCLS) data products\footnote{https://doi.org/10.48479/7epd-w356}.

Alongside a subset of generic WIMP annihilation channels, namely $b\bar{b}$, $\mu^+ \mu^-$ and $\tau^+ \tau^-$, this work aims to probe the DM candidate within the two-Higgs Doublet Model with additional Singlet Scalar (2HDM+S). 2HDM+S is a particle physics model introduced as an explanation for anomalies in the multi-lepton final states and a distortion of the Higgs transverse momentum spectrum that have been documented in Run 1 and 2 data from the Large Hadron Collider (LHC) experiments, A Toroidal LHC Apparatus (ATLAS)~\cite{aad2008atlas} and the Compact Muon Solenoid (CMS)~\cite{linden2008cms}. This candidate is of particular interest as the conjectured mass range of this candidate overlaps with that of the astrophysically motivated DM models for the Payload for Antimatter Exploration and Light-nuclei Astrophysics (PAMELA)~\cite{picozza2007pamela} anti-particle excesses and the Galactic center gamma-ray excess observed by Fermi-LAT.  

 We produce constraints on the DM parameter space for annihilation through four channels for five galaxy clusters. The most constraining results, for Abell 4038, assuming an NFW density profile, are comparable to the most stringent previous results for generic WIMPS, excluding annihilation through bottom quarks for WIMP masses below 1000 GeV. Additionally, we have begun to probe the 2HDM+S parameter space beyond the forecast levels for dwarf galaxies, and the results show that the 2HDM+S remains a viable explanation for various astrophysical excesses.

This paper is structured to introduce the 2HDM+S particle physics model in Section~\ref{section:2hdm+s}. The formalism of radio emissions from WIMP annihilation is then reviewed in Section~\ref{section:radiotheory}. In Section~\ref{section:MGCLS} we present the galaxy cluster sample investigated, with the results presented in Section~\ref{section:Results}. We then present our conclusions in Section~\ref{conclusion}. Throughout this work we assume a $\Lambda$-CDM cosmology, with $\Omega_{\mathrm{m}}= 0.3089$, $\Omega_{\Lambda}=0.6911$ and the Hubble constant $H_0=67.74\, \mathrm{km} \, \mathrm{s}^{-1} \, \mathrm{Mpc}^{-1}$ ~\cite{ade2016planck}.

\section{Two Higgs Doublet Model with singlet scalar}\label{section:2hdm+s}
Despite the success of the Standard Model (SM) of particle physics in its description of the subatomic world, it is incomplete. The open questions include the composition of dark matter and dark energy and the asymmetry between matter and anti-matter. The existence of Beyond Standard Model (BSM) physics is alluded to by anomalies at particle colliders, for example, new resonances, non-resonant states, or any deviations in the final states from the precise predictions of the SM~\cite{virdee2016beyond}. Data from run 1 and 2 at ATLAS~\cite{aad2008atlas} and CMS~\cite{linden2008cms} have reported various anomalies in multi-lepton final states as well as a distortion of the Higgs transverse momentum spectrum~\cite{von2018multi,von2020anatomy,hernandez2021anomalous}. 

The two-Higgs Doublet Model with an additional singlet scalar (2HDM+S) \cite{von2019constraints} is a BSM particle physics model that was introduced as an explanation for the various anomalies. The model includes a heavy scalar boson, H with m$_\mathrm{H}=270$ GeV, a scalar mediator, S, and a dark matter candidate $\chi$. Scalar S acts as a mediator to the dark sector through the decay chains H $\rightarrow$ hS, SS and S$\rightarrow \chi\chi$, where h represents the Higgs boson \cite{von2016phenomenological}. An implication of this model in the production of multiple leptons as final products of the above decay chains. Statistically compelling excesses for opposite and same sign di-leptons as well as three lepton channels with and without b-tagged jets were reported by \cite{von2019emergence,von2020anatomy,hernandez2021anomalous}. The multi-lepton excesses were examined in \cite{von2018multi}, and the best fit to the data was obtained when $m_\mathrm{S}= 150 \pm 5$ GeV. Evidence for the production of a candidate for S with mass $151$ GeV was obtained by combining sideband data from SM Higgs searches~\cite{crivellin2021accumulating}. If all decay channels are included, a global significance of 4.8$\sigma$ was reported for the mass range (130-160 GeV) required to explain the anomalies~\cite{crivellin2021accumulating}.

The potential of 2HDM+S~\cite{von2019constraints} is expressed by: 
	\begin{multline}
		V(\Phi_1,\Phi_2,\Phi_S)=m_{11}^2|\Phi_1|^2+m_{22}^2|\Phi_2|^2-m_{12}^2\left( \Phi_1^\dag \Phi_2 +\mathrm{h.c.}\right) \\ +\frac{\lambda_1}{2}\left(\Phi_1^\dag\Phi_1\right)^2  +\frac{\lambda_2}{2}\left(\Phi_2^\dag\Phi_2\right)^2 \\  +\lambda_3\left(\Phi_1^\dag\Phi_1\right) \left(\Phi_2^\dag\Phi_2\right) +\lambda_4\left(\Phi_1^\dag\Phi_2\right) \left(\Phi_2^\dag\Phi_1\right)\\+\frac{\lambda_5}{2}\left[\left(\Phi_1^\dag\Phi_2 \right)^2+\mathrm{h.c.} \right] +\frac{1}{2} m_S^2 \Phi_S^2 \\+\frac{\lambda_6}{8} \Phi_S^4+\frac{\lambda_7}{2}\left(\Phi_1^\dag \Phi_1\right)\Phi_S^2 +\frac{\lambda_8}{2}\left(\Phi_2^\dag \Phi_2\right)\Phi_S^2\; .
	\end{multline}
 
The fields $\Phi_1$ and $\Phi_2$ are the $SU(2)_L$ Higgs doublets, and terms with subscript S are contributions from the singlet field. A more in-depth presentation of the formalism of this model, the interaction Lagrangians and the constraints of the parameter space can be found in \cite{von2016phenomenological, von2019constraints}. The interaction Lagrangian between S and a dark matter candidate, $\chi$, with spin 0, 1/2, or 1 is expressed by Beck et al~\cite{beck2021connecting} as:
\begin{equation}
	\mathcal{L}_{0} = \frac{1}{2}m_{\chi} g_{\chi}^S \chi\chi S \; ,
\end{equation}

\begin{equation}
		\mathcal{L}_{1/2}= \overline{\chi}( g_\chi^S +i g_\chi^P \gamma_5)\chi S \; ,
\end{equation}
and
\begin{equation}
	\mathcal{L}_{1}= g_{\chi}^S \chi^{\mu} \chi_{\mu} S \; ,
\end{equation}
respectively. The factor $g_{\chi}^S$ describes the strength of the scalar coupling between S and the dark matter candidate, $g_\chi^P$ is the strength of the pseudo-scalar coupling to S, and $m_\chi$ is the mass of the dark matter candidate.

 WIMP-like DM candidates can annihilate into pairs of quarks, leptons, Higgs, and other bosons. The hadronization and further decay of these particles then lead to the electrons/positrons that induce the synchrotron radiation investigated in this work. When studying 2HDM+S the simplest interactions that can be considered are  $\chi \overline{\chi} \rightarrow S \rightarrow X $ and $\chi \overline{\chi} \rightarrow S \rightarrow H S/h \rightarrow X $, where X represents the SM products of the annihilation, $e^-/e^+$ for our purposes. The lowest DM mass that can be probed through these reactions is that which can produce S, or H and h together respectively. For the former, the lower limit is approximately 75 GeV, while for the latter the lower limit is approximately 200 GeV. 

 The per annihilation spectra of these two processes have been computed through Monte Carlo simulations by Beck et al~\cite{beck2021connecting}. A notable result from \cite{beck2021connecting} is that the particle yield functions for 2HDM+S do not vary significantly for the choice of spin of the dark matter candidate, when spin 0, 1/2, and 1 were investigated. As such, we limit our investigation to the spin-0 candidate.

\section{Radio emission from dark matter}\label{section:radiotheory}
In this section, we detail the procedure that is followed in order to produce a model of the dark matter radio emission. The basic recipe of this procedure is: 
\begin{itemize}
    \item Determine the source function of the electrons. This is proportional to the square of the dark matter density as well as the WIMP mass, according to the following relation 
    $ Q_e(r,E)= \langle \sigma v \rangle  \sum_{f} \frac{dN^f}{dE} \mathcal{B}_f \frac{\rho_{\chi}^2}{2 \mathrm{M}_{\chi}}$.
    \item Solve for the equilibrium distribution of the electrons $\phi(r,E)$, by solving the diffusion-loss equation. 
    \item The synchrotron emissivity is then found by integrating the product of the electron equilibrium distribution and the synchrotron power, $j_{\mathrm{sync}}(\nu,r,z)= 2 \int_{m_e}^{M_\chi} dE \ \psi(E,r) P_{\mathrm{sync}}(\nu,E,r,z)$.   
\end{itemize}
The following subsections provide further details and discussions on each of these steps, where the modeling of the dark matter signal is detailed in \ref{synch}, the electron equilibrium distribution can be obtained via Green's method (outlined in \ref{Greens}) or an operator splitting method (\ref{OSmethod}).
\subsection{Synchrotron emission}\label{synch}
\begin{figure}
    \centering
    \includegraphics[width=0.95\linewidth]{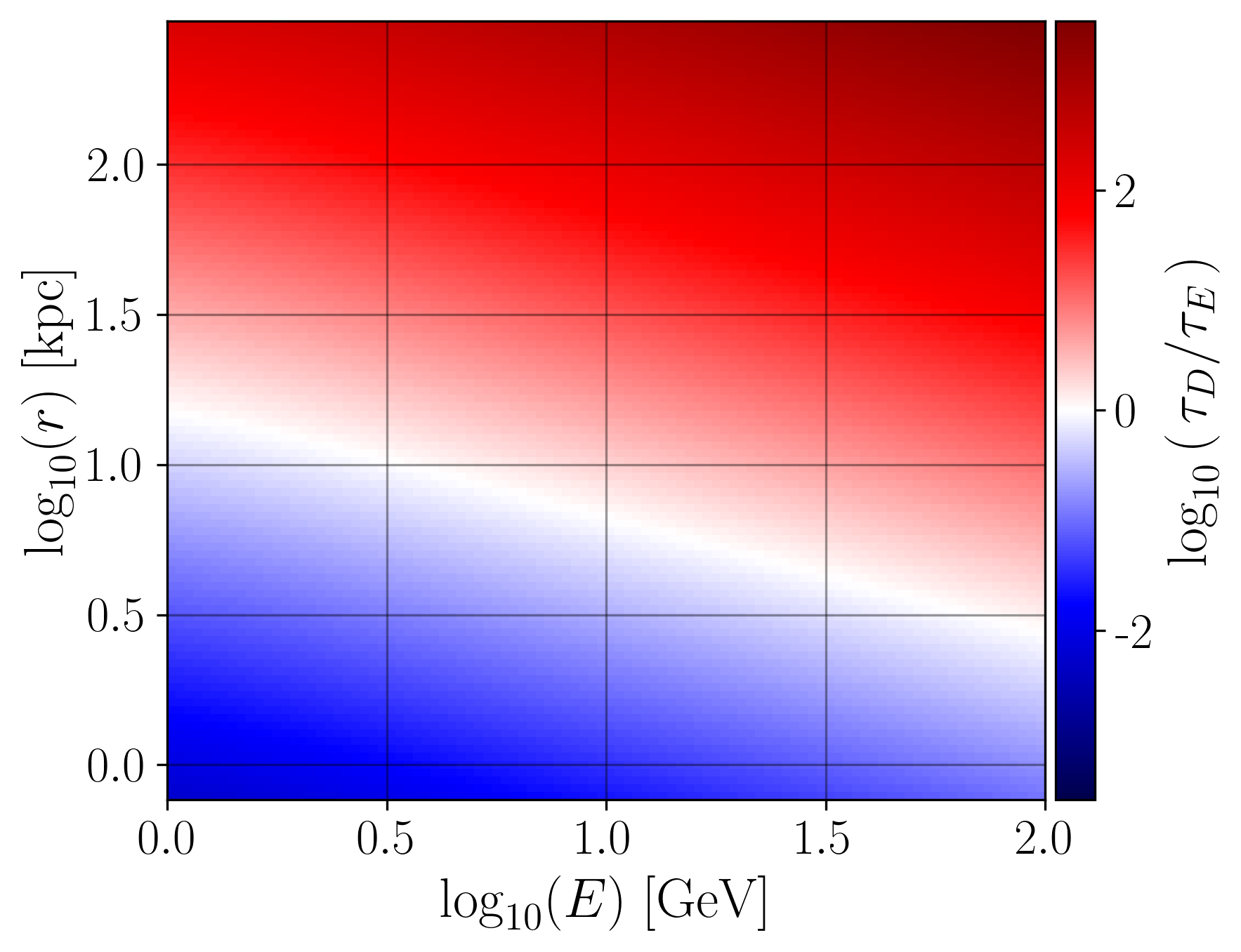}
    \caption{A two-dimensional plot indicating the ratio of the characteristic timescale of diffusion ($\tau_D$) to the timescale of energy losses ($\tau_E$) for a range of electron energies and radial position from the center of the dark matter halo. The radius has been restricted to physically relevant scales, where the minimum scale is limited by the size of a pixel on the MeerKAT FITs images and the maximum scale is the scale radius of the dark matter halo. Abell 4038 has been used as an illustrative example.    }
    \label{fig:ratio_of_timescalesssssssss}
\end{figure}

Radio emission can be a product of DM annihilation when relativistic electrons and positrons are products of the process. The interaction of these particles with the magnetized environment within the DM halo then produces synchrotron emission. Positrons and electrons will be continuously injected into the halo environment from each DM annihilation. This is typically described through a source function, which can be expressed as: 
\begin{equation}
    Q_e(r,E)= \langle \sigma v \rangle  \sum_{f} \frac{dN^f}{dE} \mathcal{B}_f \mathcal{N}_\chi(r)\; ,
\end{equation}
where $\langle \sigma v \rangle$ is the velocity averaged annihilation cross section for dark matter, $\mathcal{B}_f$ is the branching ratio to state f, and the particle production spectra for each state is $\frac{dN^f}{dE}$. For general WIMP channels we make use of $\frac{dN^f}{dE}$ from \cite{Cirelli_2011,Ciafaloni_2011}, while for 2HDM+S we use results from \cite{beck2021connecting}. The factor $\mathcal{N}_\chi(r)$ details the DM pair density, which can be further expressed as $\mathcal{N}_\chi(r)= \frac{\rho_{\chi}^2}{2 M_{\chi}^2}$. The spectra of injected electrons will evolve with time, as the particles diffuse and lose energy. It is therefore necessary to solve for the equilibrium distributions before determining the synchrotron emission. In many works that consider galaxy clusters, e.g. \cite{chan2019fitting, chan2020constraining,chan2020,chan2020possible}, the effects of diffusion are neglected due to the cooling time of the electrons being much smaller than the diffusion scale of galaxy clusters. This simplifies the diffusion-loss equation that is then solved. While the effects of diffusion on large scales are minimal it is in principle more robust to include diffusion when solving for the equilibrium distributions of the electrons. In addition, relatively small scales within the target clusters are considered in this work, thus necessitating the inclusion of the diffusion effects. In Figure \ref{fig:ratio_of_timescalesssssssss} the timescales of diffusion and energy losses are depicted for a range of radii within the cluster and electron energies. Here it can be seen that within the scale radius of the dark matter halo, diffusion occurs quickly and therefore contributes significantly to the diffusion-loss equation. It is therefore robust to include diffusion when solving for the equilibrium distribution of electrons within our regions of interest.

Within this work, two methods are utilized to obtain the required equilibrium electron distributions, which we have outlined in sections \ref{Greens} and \ref{OSmethod}. Synchrotron emission is produced when the high energy electrons, of energy E, interact with the magnetic field, of strength B. The power of this emission is dependent on both the observed frequency $\nu$ and the redshift, z, of the source. It is expressed as~\cite{longair2010high}: 
\begin{equation}
	P_{\mathrm{sync}}(\nu,E,r,z)= \int_{0}^{\pi} d\theta \ \frac{\sin^2 \theta}{2} 2 \pi \sqrt{3}r_e m_e c \nu_g F_{\mathrm{sync}} \left( \frac{\kappa}{\sin \theta} \right) \; ,
\end{equation} 

with $m_e$ as the mass of an electron, $r_e= \frac{e^2}{m_ec^2}$ the classical radius of an electron, and the non-relativistic gyro-frequency $\nu_g = \frac{cB}{2\pi m_e c}$. The parameter $\kappa$ is defined as:  
\begin{equation}
	\kappa= \frac{2\nu (1+z)}{3 \nu_0 \gamma} \left( 1+ \left(\frac{\nu_p \gamma}{\nu (1+z)} \right)^2 \right)^{3/2} \; ,
\end{equation}
here $\nu_p$ is the plasma frequency, which is directly dependent on the electron density of the environment being modelled. The function $F_{\mathrm{sync}}$ gives the synchrotron kernel, with form 
\begin{equation}
	F_{\mathrm{sync}}(x) \approx 1.25x^{\frac{1}{3}} \exp^{-x}(648+x^2)^{\frac{1}{12}}\; .
\end{equation}

The synchrotron emissivity at a radius r within the dark matter halo is then found be to 
\begin{equation}
	j_{\mathrm{sync}}(\nu,r,z)= 2 \int_{m_e}^{M_\chi} dE \ \psi(E,r) P_{\mathrm{sync}}(\nu,E,r,z) \; ,
\end{equation}
where the factor of $2$ comes from adding the $\psi$'s for electrons and positrons. The emissivity is then used to calculate the flux density and the azimuthally averaged surface brightness that would be observed. These quantities have the form~\cite{beck2019radio}  
\begin{equation} \label{flux}
	S_{\mathrm{sync}} (\nu,z) = \int_{0}^{r} d^3r' \ \frac{j_{\mathrm{sync}}(\nu,r,z)}{4\pi D_L^2}\; ,
\end{equation}
with $D_L$ the luminosity distance to the target, and 
\begin{equation} \label{brightness}
	I_{\mathrm{sync}}(\nu, \Theta, \Delta \Omega,z) = \int_{\Delta\Omega} d \Omega \int_{\mathrm{l.o.s}} d l \ \frac{j_{\mathrm{sync}}(\nu,r,z)}{4 \pi \Delta \Omega} \; ,
\end{equation}
respectively. Note that, in Eq.~(\ref{brightness}), $\Delta \Omega$ represents the angular binning scale and $\Theta$ is the angular distance of the line of sight from the cluster centre.

 The results of this modeling procedure can then be compared to measurements of astrophysical sources to probe the DM parameter space. A common way to go about this is to compare the integrated fluxes of a region of interest.

\subsection{Diffusion of electrons: Green's functions }\label{Greens}
The required electron distribution $\psi(E, \mathbf{x})$ is obtained by solving the diffusion-loss equation under the assumption of vanishing time derivatives, 
\begin{equation}\label{equation:equlibrium} 
		\nabla \left(D(E, \mathbf x) \nabla \psi \right) + \frac{\partial}{\partial E} \left(b(E, \mathbf x ) \psi \right) 
  + Q_\mathrm{e}(E,\mathbf x) = 0 \; .
	\end{equation}
In the above $D(E, \mathbf x)$ is the diffusion function and $b(E, \mathbf x)$ is the energy loss function. In order to facilitate the solution of this equation we utilize a Green's function method. This requires that the diffusion and loss functions have no spatial dependence. We define the diffusion function under the assumption of Kolmogorov turbulence via~\cite{blasi2001non}: 
\begin{equation}
D(E) = \ 3\times 10^{28} \left(\frac{d_{0,\mathrm{kpc}}^2 E_{\mathrm{GeV}}}{\overline{B}_{\mu\mathrm{G}}}\right)^{1/3}  \ \mathrm{cm}^2 \ \mathrm{s}^{-1}   \; ,
\end{equation}
where $d_0$ is the coherence length of the magnetic field, $\overline{B}$ is the average magnetic field, $d_{0,\mathrm{kpc}} = \left(\frac{d_0}{1 \; \mbox{kpc}}\right)$, $\overline{B}_{\mu\mathrm{G}} = \left(\frac{\overline{B}}{1 \; \mu \mathrm{G}}\right)$, and $E_{\mathrm{GeV}} = \left(\frac{E}{1 \; \mathrm{GeV}}\right)$. Although there is an uncertainty in the choice of diffusion coefficient for extragalactic targets, the final results are robust to variations in this value \cite{sarkis2023radio}. Additionally, Beck et al \cite{beck2022galaxy} investigated the effect of varying diffusion assumptions. By comparing results obtained under the assumption of Kolmogorov turbulence to those obtained assuming Bohmain diffusion, which is given by \cite{blasi2001non}:
\begin{equation}
    D_{\mathrm{Bohm}}(E)= 3.3\times10^{22} \ \frac{E_{\mathrm{GeV}}}{\overline{B}_{\mu\mathrm{G}}} \ \mathrm{cm}^2 \ \mathrm{s}^{-1} \; ,
\end{equation}
the authors found that for the different diffusion scenarios, there is a relatively small difference in the predicted surface brightness at small radii only. As such in this work, the results presented assume Kolmogorov turbulence and a diffusion constant $D_0= 3\times10^{28}\ \mathrm{cm}^2 \ \mathrm{s}^{-1}$.

The energy loss function is given by:  
\begin{align}
b(E) & = b_{\mathrm{IC}} E_{\mathrm{GeV}}^2 + b_{\mathrm{sync}} E_{\mathrm{GeV}}^2 \overline{B}_{\mu\mathrm{G}}^2 \;\nonumber\\ & + b_{\mathrm{Coul}} \overline{n}_{\mathrm{cm}3} \left(\log\left(\frac{\gamma}{\overline{n}_{\mathrm{cm}3}}\right)+73\right)\nonumber \\& + b_{\mathrm{brem}} \overline{n}_{\mathrm{cm}3} \gamma \left(\log\left(\gamma\right) + 0.36 \right)\;,
\end{align}
as in~\cite{egorov2022updated}. In this expression $ \gamma =\frac{E}{m_e c^2}$, the average gas density is $\overline{n}$,  $\overline{n}_{\mathrm{cm}3} = \left(\frac{\overline{n}}{1 \; \mbox{cm}^{-3}}\right)$, $ \overline{B}$ is the average magnetic field and $\overline{B}_{\mu\mathrm{G}} = \left( \frac{\overline{B}}{1 \, \mu \mathrm{G}} \right)$. Each term represents the energy loss rate of the given process, these being Inverse Compton scattering, synchrotron emission, Coulomb scattering, and bremsstrahlung, in the order they appear. The values of the numerical coefficients are $0.25\times10^{-16}( \mathrm{1+z})^4$ (for scattering with CMB photons), $0.0254\times 10^{-16}$, $7.6 \times10^{-18}$ and $7.1\times10^{-20}$ respectively in units of $\mathrm{GeV \, s}^{-1}$. We note that the utilization of a Green's function to solve for the equilibrium electron distributions requires that the diffusion and loss functions have no spatial dependence. This necessitates the use of an average for the gas density and magnetic field. These radial averaged values are calculated within the scale radius of the dark matter halo, to ensure that they are reflective of the environment in which a majority of annihilations will occur ~\cite{beck2022galaxy}. For comparison the operator splitting method outlined in section \ref{OSmethod} retains the spatial dependence of these functions.  

The equilibrium solutions to Eq.~(\ref{equation:equlibrium}), assuming spherical symmetry, are given by \cite{baltz1998positron,baltz2004diffuse}:
\begin{equation}
	\psi(r,E)= \frac{1}{b(E)}\int_{E}^{M_{\chi}} dE^\prime \ G(r,E,E')Q(r,E') \; .
\end{equation}

The Green's function has the form 
\begin{equation}
	G(r,E,E')= \frac{1}{\sqrt{4 \pi \Delta v}} \sum_{n=-\infty}^{\infty} (-1)^n \int_{0}^{r_h} dr^\prime \ \frac{r'}{r_n} f_{G,n} \; ,
\end{equation} 
where $r_h$ is the diffusion limit, where it is expected that $\psi$ approaches zero, and $n \in \mathbb{Z}$.
\begin{equation}
	f_{G,n}= \left( \exp\left(-\frac{(r'-r_n)^2}{4 \Delta v}\right) -\exp\left(-\frac{(r'+r_n)^2}{4 \Delta v}\right) \right) \frac{Q(r')}{Q(r)} \; ,
\end{equation}
 with internal functions expressed as~\cite{ beck2019radio}:
\begin{equation}
	\Delta v = v(E)-v(E') \; ,
\end{equation}
and
\begin{equation}
	v(E) = \int_{E}^{M_\chi} dx \ \frac{D(x)}{b(x)} \; .
\end{equation}

\subsection{Diffusion of electrons: OS method}\label{OSmethod}
This solution method is an operator splitting (OS) method ~\cite{porter2022galprop,press2007} which follows the scheme used in the public code \verb|galprop|~\cite{strong1998} and in \cite{regis2017dark}.

In this case, we take the diffusion function in galaxy clusters to be
\begin{equation}
D(E,r) = 10^{29}\left(\frac{B(r)}{B(0)}\right)^{-\frac{1}{3}}E_{\mathrm{GeV}}^{\frac{1}{3}} \ \mathrm{cm}^2 \ \mathrm{s}^{-1}  \; . \label{eq:diff-adi}
\end{equation}
This is designed to generate a similar magnitude to the Green's approach in the cluster centre. We then discretize the diffusion-loss equation according to the scheme
\begin{multline}
	\dfrac{\psi_i^{n+1}-\psi_i^n}{\Delta t} = \dfrac{\alpha_1 \psi_{i-1}^{n+1} - \alpha_2 \psi_{i}^{n+1} + \alpha_3 \psi_{i+1}^{n+1}}{2\Delta t} \\[0.5ex]
	+ \dfrac{\alpha_1 \psi_{i-1}^{n} - \alpha_2 \psi_{i}^{n} + \alpha_3 \psi_{i+1}^{n}}{ 2\Delta t} + Q_{e,i} \; ,  
\end{multline}
where $i$ are spatial or energy indices and $n$ are those for time. The $\alpha$ coefficients are determined by matching to Eq.(\ref{equation:equlibrium}) with non-zero time-derivative of $\psi$ on the right-hand side.
This can be re-arranged to form a tri-diagonal equation 
\begin{multline}\label{eqn:cn}
	-\dfrac{\alpha_1}{2}\psi^{n+1}_{i-1} + \left(1+\dfrac{\alpha_2}{2}\right)\psi^{n+1}_{i} - \dfrac{\alpha_3}{2}\psi^{n+1}_{i+1} \\
	= Q_{e,i}\Delta t + \dfrac{\alpha_1}{2}\psi^{n}_{i-1} + \left(1- \dfrac{\alpha_2}{2}\right)\psi^{n}_{i} + \dfrac{\alpha_3}{2}\psi^{n}_{i+1} \; .
\end{multline}
matching the form $A\psi^{n+1} = B\psi^{n} + Q_e$ where $A$ and $B$ are tri-diagonal matrices.

\subsubsection{Operator splitting}
The time derivative is then split into two operators, one depending on energy and the other on radius. Each of these will be acted independently on $\psi$ sequentially to produce the full update from time $n$ to $n+1$. We then transform the variables $E$ and $r$ to use a logarithmic scale, to better account for the range of physical scales involved, \textit{i.e.} $\tE = \log_{10}(E/E_0)$ and $\tr = \log_{10}(r/r_0)$, where $E_0$ and $r_0$ are chosen scale parameters.  We discretize the $r$ derivative term as in a generalized Crank-Nicolson scheme~\cite{press2007}
\begin{multline}
	\dfrac{1}{r^2}\pd{}{r}{}\left(r^2D\pd{\psi}{r}{}\right) \rightarrow \\
	(r_0\log(10)10^{\tr_i})^{-2}\left[\dfrac{\psi_{i+1}-\psi_{i-1}}{2\Delta \tr}\left.\left(\log(10)D	+ \pd{D}{\tr}{}\right)\right\vert_{i}\right. \\
    + \left.\dfrac{\psi_{i+1}-2\psi_{i}+ \psi_{i-1}}{\Delta \tr^2}D\vert_{i} \right] \; ,
\end{multline} 
with the energy derivative
\begin{equation}
	\pd{}{E}{}(b\psi) \rightarrow (E_0\log(10)10^{\tE_j})^{-1}\left[\dfrac{b_j\psi_{j+1}-b_j\psi_{j}}{\Delta \tE}\right] \; ,
\end{equation}
where $\Delta \tr$ and $\Delta \tE$ represent the radial and energy grid spacings,  and the time superscripts have been suppressed as they are all identical. Note that in the case of the energy derivative we use only forward differencing, as energy is only lost in the problem, not gained. The updating scheme is then given by
\begin{align}
    \psi^{n+1/2} &= \Psi_{\tE}(\psi^{n}) \; , \\
    \psi^{n+1} &= \Psi_{\tr}(\psi^{n+1/2}) \; ,
\end{align}
where $\Psi_{\tE}$, and $\Psi_{\tr}$ are the partial updating operators:
\begin{equation}\label{eqn:alpha_r}
	\Psi_{\tr}:
	\begin{cases}
		\dfrac{\alpha_1}{\Delta t} &= C_{\tr}^{-2}\left.\left(-\dfrac{\ln(10)D+\frac{\partial D}{\partial\tr}}{2\Delta\tr} + \dfrac{D}{\Delta\tr^2}\right)\right\vert_{i}\;,\\[3ex]
		\dfrac{\alpha_2}{\Delta t} &= C_{\tr}^{-2}\left.\left(\dfrac{2D}{\Delta\tr^2}\right)\right\vert_{i} \; ,\\[3ex]
		\dfrac{\alpha_3}{\Delta t} &= C_{\tr}^{-2}\left.\left(\dfrac{\ln(10)D+\frac{\partial D}{\partial\tr}}{2\Delta\tr} + \dfrac{D}{\Delta\tr^2}\right)\right\vert_{i} \; ,
	\end{cases}
\end{equation}
and
\begin{equation}\label{eqn:alpha_E}
	\Psi_{\tE}:
	\begin{cases}
		\dfrac{\alpha_1}{\Delta t} &= 0 \; ,\\[3ex]
		\dfrac{\alpha_2}{\Delta t} &= C_{\tE}^{-1}\dfrac{b_j}{\Delta E}\; ,\\[3ex]
		\dfrac{\alpha_3}{\Delta t} &= C_{\tE}^{-1}\dfrac{b_{j+1}}{\Delta E} \; .
	\end{cases}
\end{equation}
Here $C_{\tr}=(r_0\log(10)10^{\tr_i})$ and $C_{\tE} = (E_0\log(10)10^{\tE_j})$. Note that determining $\psi^{n+1/2}$ and $\psi^{n+1}$ each require solving a tri-diagonal system of equations.

\subsubsection{Boundary conditions}
The final aspect of the solution is to specify boundary conditions. These are taken to be
\begin{align}
	\psi &= 0 \; , \qquad \tr = \tr_{\mathrm{max}} \; ,\\
	\pd{\psi}{\tr}{} &= 0 \; , \qquad \tr = \tr_{\mathrm{min}} \; .
\end{align}
These conditions then imply
\begin{equation}
	\Psi_{\tr=\tr_{\mathrm{min}}}:
	\begin{cases}
		\dfrac{\alpha_1}{\Delta t} &= 0 \; ,\\[3ex]
		\dfrac{\alpha_2}{\Delta t} &= C_{\tr}^{-2}\left.\left(\dfrac{4D}{\Delta\tr^2}\right)\right\vert_{i} \; ,\\[3ex]
		\dfrac{\alpha_3}{\Delta t} &= C_{\tr}^{-2}\left.\left(\dfrac{4D}{\Delta\tr^2}\right)\right\vert_{i} \; .
	\end{cases}
\end{equation}

\subsubsection{Convergence and time-scales}
To determine convergence we need to specify the loss and diffusion time-scales as
\begin{equation}\label{eqn:loss_ts}
	\tau_{\mathrm{loss}} = \dfrac{\tE}{b(\tE,\tr)} \; ,
\end{equation}
and
\begin{equation}\label{eqn:diff_ts}
	\tau_{D} = \dfrac{\tr_{\mathrm{min}}^2}{D(\tE,\tr)} \; .
\end{equation}
These can be compared to the time-scale on which the solution changes 
\begin{equation}\label{eqn:psi_ts}
	\tau_{\psi} = \frac{\psi}{\left\vert\left(\dfrac{\psi^{n+1} - \psi^n}{\Delta t}\right)^{-1}\right\vert} \; .
\end{equation}
When $\tau_{\psi}$ is less than both loss and diffusion time-scales at all $r$ and $E$ the solution is considered to have converged. 

In practice we adopt the same accelerated convergence as \verb|galprop|~\cite{strong1998}. This involves starting at a large time-step value $\sim 10^9$ years performing a minimum of 100 steps (we require the solution's fractional change per time-step has fallen to $10^{-5}$) and then reducing the time-step by a factor of 2. This continues till we reach a minimum scale $\sim 10$ years. Final convergence is then determined by comparison of time-scales and the requirement that the solution's fractional change per time-step has fallen to $10^{-3}$.  

\begin{figure}
    \centering
    \includegraphics[width=0.9\linewidth]{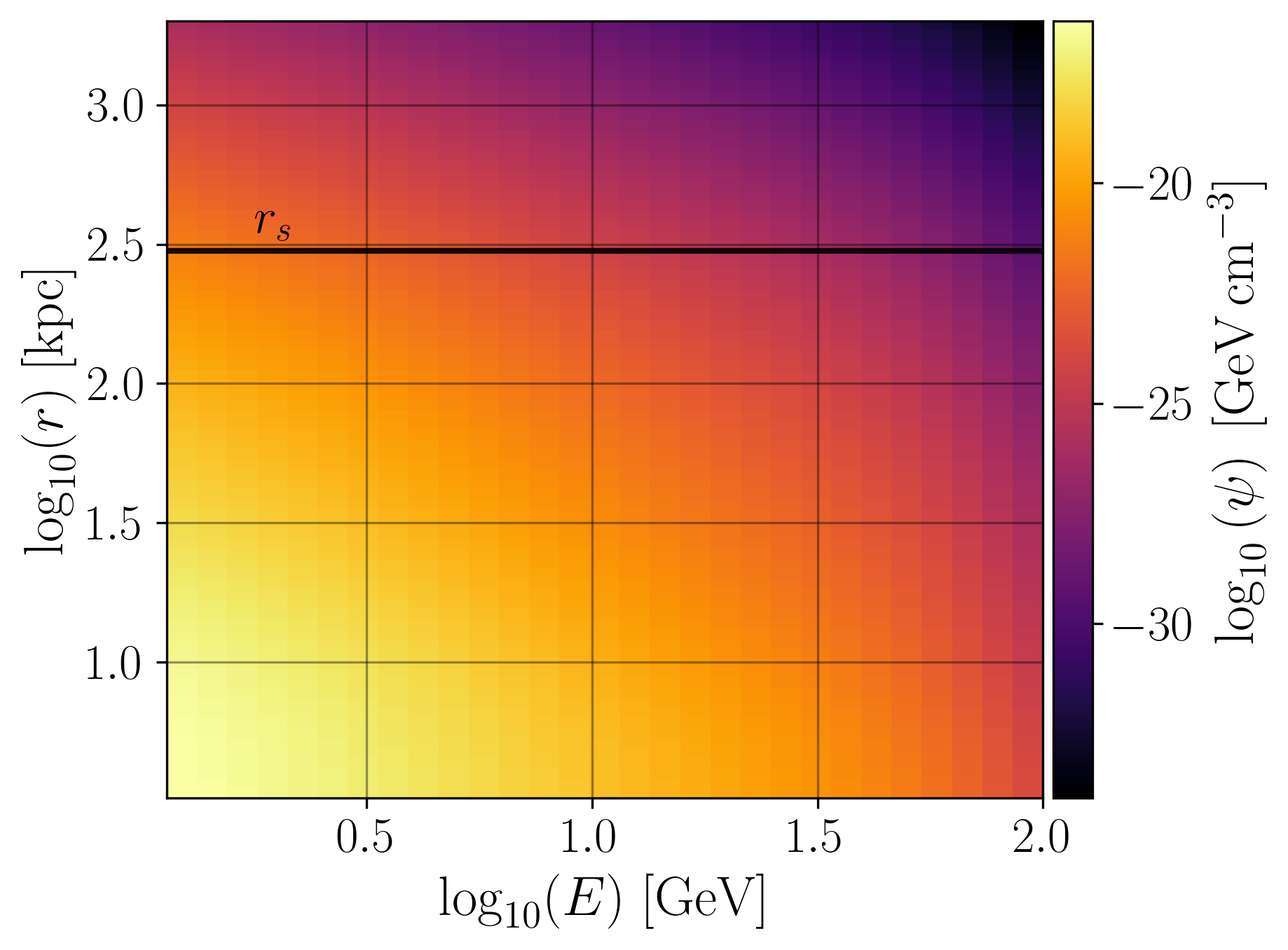}
    \caption{Log-log plot of the equilibrium electron distribution as a function of the electron energy and radial position, where we have used Abell 4038 as an illustrative example. The solid black line indicates the scale radius of the dark matter halo of Abell 4038.}
    \label{fig:electrondist}
\end{figure}

In Figure \ref{fig:electrondist} we show the equilibrium electron distribution as a function of both electron energy and radial position for Abell 4038. This plot shows that the electrons are concentrated toward the center of the cluster and fall off as the radial distance is increased.  

\subsection{Boost Factor}
 The above modeling procedure for predicting the synchrotron emission assumes a smooth DM profile, and it is reasonable to expect that the presence of sub-structure will affect the strength of the annihilation signal. Molin$\acute{e}$ et al 2017 \cite{moline2017characterization} investigate the effect of substructure on the annihilation signal. The authors present a parametric equation for the boost factor as a function of host halo mass. The boost is given by 
 
 \begin{equation}
 	\log {B}_\mathrm{sub}(M)= \sum_{i=0}^{5} b_i \left[ \log \left(\frac{M}{M_{\odot}}\right) \right]^i \; ,
 \end{equation} 
 where the best-fit parameters $b_i$ are presented for two values of the slope of the subhalo mass function, $ \alpha =1.9$ and $\alpha=2$. In this work, we utilize the parametric equation with $\alpha=2$ to calculate the full boost factor within the virial radius, denoted by $f_\mathrm{boost}$. Here, elaboration is required for the use of the term `full boost'. It must be noted that the authors formulated the expression for this factor with primarily $\gamma-$ray annihilation signals in mind. On the contrary, synchrotron emission cannot benefit from this enhancement in the signal. This is due to the fact that sub-halos are more common on the outskirts of the host halo \cite{jiang2017statistics}, where magnetic fields are weaker. Thus, it is necessary to scale the boost to the regions where synchrotron emission is more likely. This is achieved through the consideration of the spatial distribution of the substructure within the host halo, as well as the magnetic field distribution. This leads to the scaled boost taking the form presented in Beck et al 2022 \cite{beck2022galaxy}:
 
 \begin{equation} \label{boost}
 	\mathcal{B}(R)= 4 \pi \int_{0}^{R} dr\, r^2 f_\mathrm{boost} \left( \frac{B(r)}{B_0} \right) \tilde{\rho}_\mathrm{sub}(r) \; ,
 \end{equation}
 where $ \tilde{\rho}_\mathrm{sub}(r)$ is the sub-halo mass density normalized to 1 between 0 and the virial radius. It can be understood as the host halo mass density multiplied by a modifier function from \cite{jiang2017statistics}. The factor $f_\mathrm{boost}$ is the full boost factor described above. The sub-structure fluxes appear to exhibit a dependency on $B$ rather than $B^2$, and it is possible this is due to the effect of the energy losses \cite{beck2022galaxy}.
 
 The flux from the host halo is then multiplied by $\mathcal{B}(R)$ to account for the presence of substructure, and obtain the total expected flux from dark matter annihilation. It is noted however that the uncertainties of the sub-halo distribution will affect any constraints of dark matter properties that are deduced through comparisons of integrated fluxes.

\section{The galaxy cluster sample and modelling parameters}\label{section:MGCLS}

\subsection{Galaxy cluster sample}
The MeerKAT Galaxy Cluster Legacy Survey (MGCLS) \cite{knowles2022meerkat} consists of $\sim 1000$ hours of observations of 115 galaxy clusters in the L-band. The sample is heterogeneous, with no mass or redshift criteria applied in the selection. The clusters can be categorized as either `radio selected' or `X-ray selected'. Clusters in the radio-selected group have been previously searched for diffuse emission, and as a result, have a bias towards high-mass clusters that host radio halos and relics. The X-ray selected group consists of clusters selected from the Meta-Catalogue of X-ray-detected Clusters (MCXC) \cite{piffaretti2011mcxc}, chosen with the intention to form a sample without a bias to radio properties. Of the 115 clusters observed, 62 clusters were found to contain a form of diffuse cluster emission, many of which were previously undetected \cite{knowles2022meerkat}.

 MGCLS data release 1 (DR1)\footnote{https://doi.org/10.48479/7epd-w356} includes primary beam corrected images, referred to as enhanced data products. There are two types of advanced products available: 5-plane cubes consisting of the intensity at the reference frequency (1.28 GHz), spectral index, brightness uncertainty estimate, spectral index uncertainty, and the $\chi^2$ of the least squares fit, as well as a frequency cube of intensity images with 12 frequency planes. There are two resolutions provided in DR1, $7^{ \prime\prime}$ and $15^{\prime\prime}$. In this work the full resolution ($7^{\prime\prime}$) images are utilized for the identification of compact point sources, while the convolved images ($15^{\prime\prime}$) images are used to identify the faint diffuse emission of radio halos or mini-halos. Details of the data reduction process are available in \cite{knowles2022meerkat}.

Galaxy clusters investigated in this work are selected from the MGCLS catalogue based on the availability of the DM halo properties in the current literature. The sample includes 12 clusters with giant radio halos. Furthermore, 2 clusters with mini-halos and 3 clusters with non-detections of diffuse emission are investigated. These fainter sources are considered to probe the effects of the presence of large baryonic backgrounds on the DM constraints. The properties of the galaxy clusters investigated are listed in Table~\ref{table:1}, where the clusters with giant radio halos are Sample 1 and the rest are Sample 2. 
\subsection{Modelling parameters}
We model the DM halo density profile with a form of the Hernquist-Zhao density profile \cite{navarro1996astrophysical}, 

\begin{equation}
    \rho(r) =\frac{\rho_s}{(\frac{r}{r_s})^{\alpha}(1+ \frac{r}{r_s})^{3-\alpha}} \; , 
\end{equation}
where we consider two halo indexes, $\alpha=1$ which describes an NFW density profile, and $\alpha=0.5$ which describes a more shallowly cusped profile. Sarkis et al \cite{sarkis2023radio} find that a more cuspy profile results in stronger constraints on the annihilation cross-section. This is a consequence of a higher DM density in the central region of the DM halo. While the DM density profiles of the targets are uncertain, there is evidence that supports an NFW-like density profile in galaxy clusters \cite{beck2022galaxy,mamon2019structural,he2020constraining}. As such, we consider the shallowly cusped profile to be the upper bound for the uncertainty in the structure of the halo profile. 

\begin{figure}
    \centering
    \includegraphics[width=0.98\linewidth]{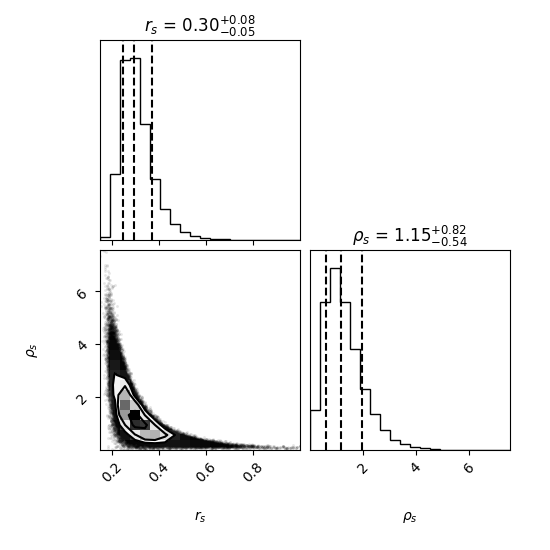}
    \caption{The corner plot of the uncertainty of $r_s$ and $\rho_s$ determined through MCMC sampling with the use of the \texttt{emcee} package for Abell 4038.}
    \label{fig: A4038corner}
\end{figure}

Our method to determine halo parameters relies on two inputs: a concentration $c_\delta$ and a mass $M_\delta$. These are linked via a contrast $\delta$ as follows:
\begin{align}
    r_\delta & = \left(\frac{3 M_\delta}{4 \pi \rho_c \delta}\right)^{1/3} \; , \\
    c_\delta & = \frac{r_\delta}{r_s} \; ,
\end{align}
where $\rho_c$ is the critical density of the universe. There are 3 common $\delta$ values: 200, 500, and 
\begin{align}
    \delta_\mathrm{vir} & = 18 \pi^2 - 80 x - 39 x^2\; , \\
    x & = 1 - \frac{1}{1+\frac{1-\Omega_{m,0}}{\Omega_{m,0}(1+z)^3}} \; .
\end{align}
If we do not possess a $c_{\delta}$ value, we estimate it aiming for consistency between $M_{200}$ and $c_{200}$, requiring they match the scaling relation from \cite{Prada_2012}. We then impose an uncertainty consistent with a $\Delta c_\mathrm{virial} = 1.4$ which is the intrinsic observational scatter from \cite{groener2016}. In some cases, such as A133, $c_{200}$ values are available~\cite{groener2016}, but are highly uncertain and vary strongly between X-ray and optical experiments. In these cases, we again resort to consistency with \cite{Prada_2012} (the resulting values fall on the lower end of estimates from \cite{groener2016} so constitute a conservative estimate).
With these in hand, we determine our halo parameters $r_s$ and $\rho_s$ by MCMC sampling using the \texttt{emcee} \footnote{https://github.com/dfm/emcee} package~\cite{Foreman_Mackey_2013} and determining appropriate posterior distributions. The calculated values are displayed in Table~\ref{tab:halo-params}. The corner plot of Abell 4038 is shown as an example in Figure \ref{fig: A4038corner}. 

\begin{table*}

		\caption[Cluster and dark matter halo properties. ]{Cluster and DM halo properties. (2): Redshift. (3)-(4): \textit{Sample 1}- NED cluster position, \textit{Sample 2} - MCXC central coordinates. (5): Halo mass. (6): Halo concentration. (7): Contrast ratio for mass and concentration.  (8): Diffuse emission (DE) classification as found in \cite{knowles2022meerkat}. (9) References. The $c$ values marked with an asterisk are estimated to match simulation results~\cite{Prada_2012} with intrinsic scatter from \cite{groener2016}.   }
        \label{table:1}
			\begin{tabular}{lcccccccc}
				\toprule
                     \centering
				Cluster name & z  & $\mathrm{R.A._{J2000}}$ (deg) & $\mathrm{Dec._{J2000}}$ (deg) & $M (10^{15} M_{\odot})$ & $c$ & $\delta$ & DE & References\\
				(1) & (2)&(3)&(4)&(5)&(6)&(7)&(8)&(9)\\
				\hline
    \textit{Sample 1} \\
    
				Abell 209 & 0.206&22.990&-13.576&$1.35 \pm 0.18$& $2.55\pm 1.18$ & 200 & Halo &\cite{knowles2022meerkat} \cite{klein2019weak}\\

				Abell 370 &0.375& 39.960&-1.586&$3.11 \pm 0.60$& $5.83\pm 0.84$ & 200 & Halo &\cite{knowles2022meerkat} \cite{umetsu_2011a}\cite{groener2016}\\

				Abell 545 & 0.154&83.102&-11.543&$0.76 \pm 0.07$& $3.4\pm 0.93$* & 500 & Halo & \cite{knowles2022meerkat} \cite{mantz2016weighing}\\
				
				Abell 2667& 0.230 &357.920&-26.084& $0.92 \pm 0.09$ & $3.4\pm0.93$* & 500 & Halo &  \cite{knowles2022meerkat} \cite{rossetti2017cool} \\
				
				Abell 2813  & 0.29&10.852&-20.621 &$1.241 \pm 0.243$ & $3.45\pm 1.15$ & 200 & Halo & \cite{knowles2022meerkat} \cite{klein2019weak}\\
				
				Abell S295 &0.3&41.4&-53.038 &$0.51 \pm 0.195$& $3.7 \pm 1.23$ & 200 & Halo & \cite{knowles2022meerkat} \cite{klein2019weak}\\
				
				Abell S1063 & 0.348&342.181&-44.529&$1.49 \pm 0.314$& $3.3\pm 1.1$ & 200 & Halo &    \cite{knowles2022meerkat} \cite{klein2019weak} \\

				J0303.7-7752 & 0.274&45.943&-77.869& $0.69 \pm 0.07$ & $3.4\pm0.93$*& 500 & Halo &  \cite{knowles2022meerkat} \cite{rossetti2017cool} \\

				J0528.9-3927 & 0.284&82.235&-39.463&$1.4 \pm 0.28$ & $5\pm 1.4^*$ & 200 & Halo & \cite{knowles2022meerkat} \cite{foex2017core} \\

				J0638.7-5358 & 0.233&99.694&-53.972  & $0.67 \pm 0.07$ & $3.4 \pm 0.93$* & 500 & Halo &  \cite{knowles2022meerkat} \cite{rossetti2017cool} \\
				
				J0645.4-5413 &0.167&101.372&-54.219 &$1.24 \pm 0.31$ & $3.65\pm 1.22$ & 200 & Halo &  \cite{knowles2022meerkat} \cite{klein2019weak} \\

				J1601.7-7544 &0.153 &240.445&-75.746& $0.75\pm 0.08$&$3.4\pm0.93$*& 500 & Halo & \cite{knowles2022meerkat} \cite{rossetti2017cool}\\
				\hline
    \textit{Sample 2}\\
				Abell 4038 &0.028 &356.93 &-28.1414 & $0.46 \pm 0.14$ & $6.66 \pm 1.4$* & Virial & Mini halo & \cite{knowles2022meerkat} \cite{wojtak2007importance} \cite{piffaretti2011mcxc} \\
				
				Abell 133 &0.057 &15.6754 &-21.8736 & $0.389 \pm 0.048$  & $6.61 \pm 1.4$* & Virial & Mini halo & \cite{knowles2022meerkat} \cite{zhu2021study} \cite{piffaretti2011mcxc}\\
				
				RXCJ0225.1-2928 &0.060 &36.2937 &-29.4739  & $0.096 \pm 0.004$ & $10.7 \pm 4.2$ & 500 & - & \cite{knowles2022meerkat} \cite{shakouri2016atca} \cite{foex2019substructures} \cite{piffaretti2011mcxc} \\
                 J0600.8-5835&0.037&90.2012&-58.5872&$0.043 \pm 0.004$ & $3.8 \pm 0.93$* & 500 & - & \cite{knowles2022meerkat} \cite{piffaretti2011mcxc}\\
                 J0757.7-5315&0.039&119.4437&-53.2636& $0.11 \pm 0.01$ & $3.4 \pm 0.93$* & 500 & - & \cite{knowles2022meerkat} \cite{piffaretti2011mcxc}\\
				\hline
			\end{tabular}

		\end{table*}

\begin{table*}
\caption{Calculated halo parameters. (2) The scale radius of the dark matter halo in Mpc. (3) The scale density of the dark matter halo. (4) Scaled boost factor.  The scale radius for the more shallowly cusped profile is $1.5$ times smaller than that of plain NFW and $\rho_s$ changes to ensure normalisation.}
\label{tab:halo-params}
\begin{tabular}{lccc}
\toprule
Cluster name & $r_s$ (Mpc) & $\rho_s$ ($10^{15}$ M$_\odot$ Mpc$^{-3}$) & Scaled boost \\
(1) & (2) & (3) & (4) \\
\hline

    \textit{Sample 1} \\
				Abell 209 & $0.69 ^{+0.16}_{-0.19}$& $0.49^{+0.19}_{-0.39}$ & 5.69  \\
				Abell 370 & $0.46^{+0.06}_{-0.08}$ & $2.35_{-1.04}^{+0.80}$ & 5.78 \\
				Abell 545 & $0.39^{+0.07}_{-0.11}$ & $2.03^{+0.57}_{-0.77}$ & 5.30 \\
				Abell 2667 & $0.409^{+0.07}_{-0.12} $ & $1.49^{+0.62}_{-0.83}$ & 1.92\\
				Abell 2813  & $0.56^{+0.13}_{-0.19}$ &  $0.72^{+0.35}_{-0.62}$ & 2.84\\
				Abell S295 & $0.40^{+0.10}_{-0.17}$ & $0.78^{+0.46}_{-0.84}$ & 5.58\\
				Abell S1063 & $0.6^{+0.13}_{-0.19}$ &  $0.73^{+0.34}_{-0.59}$ & 5.74\\
				J0303.7-7752 & $0.36^{+0.06}_{-0.10}$ & $1.59^{+0.65}_{-0.88}$ & 4.34\\
				J0528.9-3927 & $0.42^{+0.08}_{-0.12}$ & $1.49^{+0.66}_{-0.96}$ & 1.82\\
				J0638.7-5358 & $0.662^{+0.07}_{-0.10}$ & $1.51^{+0.63}_{-0.86}$ & 5.28\\
				J0645.4-5413 & $0.56^{+0.13}_{-0.20}$ & $0.71^{+0.35}_{-0.63}$ & 4.27 \\
				J1601.7-7544 & $0.39^{+0.07}_{-0.11}$ & $1.39^{+0.57}_{-0.78}$ & 3.06 \\
				\hline
    \textit{Sample 2}\\
Abell 4038 & $0.30^{+0.08}_{-0.05}$ & $1.15^{+0.82}_{-0.54}$ & \\
Abell 133 & $0.28^{+0.07}_{-0.05}$ & $1.23^{+0.73}_{-0.52}$ & \\
RXCJ0225.1-2928 & $0.06^{+0.03}_{-0.02}$ & $21.26^{+23.67}_{-14.08}$ &  \\
J0600.8-5835& $0.14^{+0.04}_{-0.02}$ & $1.56^{+0.80}_{-0.61}$ & \\
J0757.7-5315 & $0.21^{+0.06}_{-0.04}$ & $1.28^{+0.67}_{-0.050}$ &\\
\hline
\end{tabular}
\end{table*}
In addition, in order to model the synchrotron emission we must first model the electron distribution and the magnetic fields. For this, we require the gas density normalization factor, the scale radius, and the index of the radial profile, and similarly for the magnetic field. At present these properties were not available in the literature for many of the target clusters. For clusters where the information was unavailable the properties of the Coma cluster were scaled to the size of the target. The gas normalization factor is scaled according to: 

	\begin{equation}\label{eq:n0}
		n_0= n_\mathrm{Coma,0}\, \left(\frac{M_\mathrm{virial}}{M_\mathrm{Coma,virial}}\right)^{1/3} \; . 
	\end{equation}
	
	The gas scale, $r_\mathrm{e}$, and magnetic field scale, $r_\mathrm{B}$, are assumed to have the same value, following the behaviour displayed by Coma. This factor is scaled with the virial radius using the following relation:
	
	\begin{equation}\label{eq:re}
		r_\mathrm{e/B}= r_\mathrm{e/B,\, Coma}\, \left( \frac{r_\mathrm{virial}}{r_\mathrm{virial,\,Coma}} \right) \; .
	\end{equation}
 The properties of the Coma cluster are listed in Table \ref{table:Coma}.
	\begin{table}
		\centering
		\caption[Coma cluster properties]{Properties of the Coma cluster used in the scaling relations to obtain values for similar clusters.}
             \label{table:Coma}
		\begin{tabular}{lll}
			\toprule
			Property & Value& Reference \\
			\hline
			$B_0$ & 4.7 $\mu G$ & \cite{bonafede2010coma}\\
			$\eta$ & 0.5 & \cite{bonafede2010coma}\\
			$n_0$ & 3.49 $\times 10^{-3} \mathrm{cm}^{-3}$& \cite{chen2007statistics}\\
			$\beta$ & -0.654 & \cite{chen2007statistics}\\
			$r_e$ & 253 kpc & \cite{chen2007statistics}\\
			$M_\mathrm{vir}$& $1.24\times 10^{15} M_{\odot}$& \cite{lokas2003monthly}\\
			$r_\mathrm{vir} $& 2.7 Mpc & \cite{lokas2003monthly}\\
			\hline
		\end{tabular}
	\end{table}
For Abell 133, Abell 4038, and RXCJ0225.1-2928 properties of the gas density distribution were found in the literature.  The gas density distribution for Abell 133 is best described by a double beta profile, while single beta profiles provide more realistic indexes for Abell 4038 and RXCJ0225.1-2928. The gas density distribution parameters for these clusters are listed in Table \ref{table:electrons}. The gas density distribution parameters for J0600.8-5835 and J0757.7-5315 are scaled with the Coma parameters according to equations \ref{eq:n0} and \ref{eq:re}.  

\begin{figure}
    \centering
    \includegraphics[width=\linewidth]{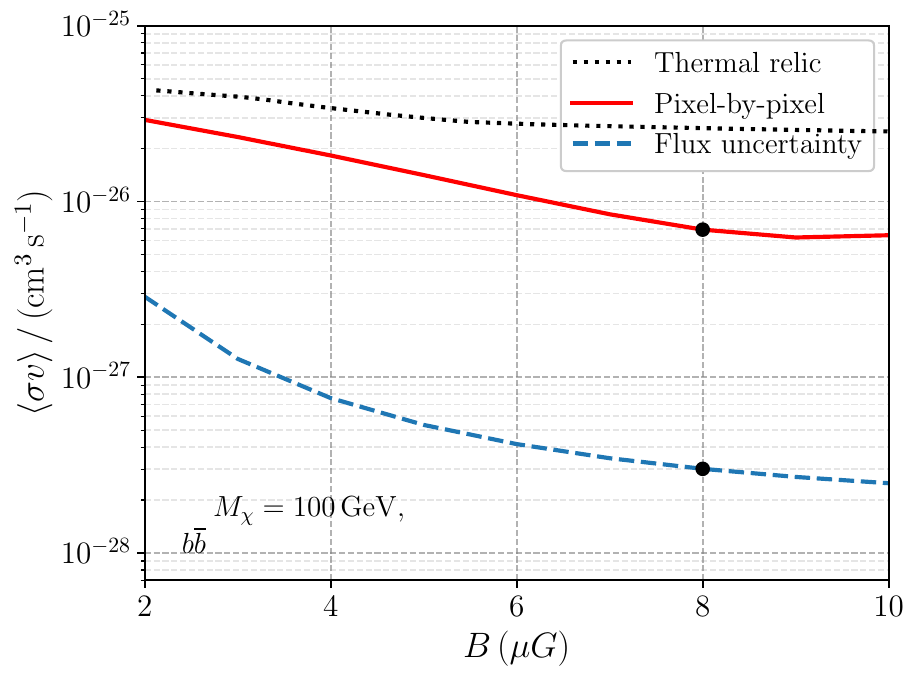}
    \caption{The impact of the magnetic field strength on the bound on the annihilation cross-section, in an example taking $M_{\chi}=100\, \mathrm{GeV}$, annihilation through $b\overline{b}$, and an NFW density profile for the median scale parameters of Abell 4038. The points show the magnetic field strength used in the analysis. }
    \label{fig:Bfield}
\end{figure}

The magnetic field strength, $B_0$, does not have a mass dependence but rather depends on the dynamical activity occurring within the cluster. Faraday rotation analyses of radio sources within or behind the cluster are a key technique used to obtain information on the cluster's magnetic field strength. Govoni et al \cite{govoni2004magnetic} find that rotation measures are extremely high in clusters that contain cool cores. The derived magnetic field strengths range from $5 \, \mu$G up to $30\, \mu$G in extreme cases. Dynamically disturbed clusters have rotation measure data that is consistent with $\mu$G central magnetic field strengths, with values $\sim 2- 8 \, \mu$G. A more recent study by Osinga et al \cite{osinga2022detection} utilizes the depolarisation of radio sources within and behind clusters to probe magnetic field parameters. Through a statistical analysis, the authors found that the best-fitting models have a central magnetic field strength $5-10 \, \mu$G. Due to insufficient detections within the core regions, no significant difference between cool-core and non-cool-cores was observed. We take a conservative estimation of $5\, \mu$G  and $8\, \mu$G for the magnetic field normalization for disturbed clusters and cool-core clusters respectively. This central value is a source of uncertainty and as such we consider the potential effects of the central magnetic field strength on the upper limits of the annihilation cross-section. An illustrative example of the effects of $B_0$ on the upper limits for Abell 4038 is shown in Figure \ref{fig:Bfield}. We take $M_{\chi}=100$ GeV, annihilation through bottom quarks, and an NFW DM density profile for Abell 4038. The bounds on the annihilation cross-section scales with the inverse square of the magnetic field strength for small values and flattens with larger values. The pixel-by-pixel method shows less variation in $\langle \sigma v \rangle$, across the considered range. This is likely due to the OS method incorporating the spatial dependence of the magnetic field, rather than taking an average value, and therefore reducing the uncertainty of the results due to the modeling parameters.

We take $2\mu$G as the lower limit for the central magnetic field, following values found for other galaxy clusters in literature e.g.  \cite{stuardi2021intracluster,govoni2006intracluster,guidetti2008intracluster}. We find that for the cool-core clusters, the limits are weakened by an order of magnitude for the flux uncertainty method and roughly a factor of 4 for the pixel-by-pixel method. For the non-cool-cored clusters, the limits are weakened by a factor of 7 and 4 for the flux uncertainty method and the pixel-by-pixel methods respectively when $B_0$ is reduced from 5 $\mu$G to 2 $\mu$G.

As many magnetic field parameters are currently unavailable we utilize general scaling relations that have been noted by Faraday rotation measure studies performed on clusters of galaxies \cite{murgia2004magnetic,govoni2006intracluster,vacca2012intracluster}. A common choice is to assume that the magnetic field distribution follows that of the gas density through \begin{equation}
		B(r)=B_0 \left( \frac{n_\mathrm{e}(r)}{n_0}\right)^{\eta} \; ,
	\end{equation} 
	predicted through both cosmological simulations and the comparison between thermal and radio brightness profiles \cite{feretti2012clusters}. The choice of $\eta=0.5$ corresponds to a magnetic field whose energy density decreases with radius in the same way as the gas density \cite{murgia2004magnetic}.

For Sample 1 the DM upper limits are determined by a 2$\sigma$ exclusion of the DM signal from the measured diffuse flux of the radio halos present. Due to the much larger physical scales investigated in this sample, the effects of varying central positions are negligible.   For Sample 2 the DM signal is injected into the MeerKAT surface brightness maps. The signal would ideally be centered on the center of the DM distribution. This position is typically estimated by the center of the galaxy cluster. However, the various measures of the cluster center (e.g. the peak and centroid of the X-ray luminosity, the position of the brightest central galaxy (BCG), the peak of the density and the minimum of the gravitational potential) are offset in galaxy clusters that are not in complete equilibrium. Here we consider the DM profile centered on the X-ray peak value from the Meta-catalogue of X-ray detected clusters of galaxies (MCXC) \cite{piffaretti2011mcxc}. This is a good indicator of the cluster center in relaxed clusters \cite{seppi2023offset,cui2016does}. The effects of different central positions will be discussed in Section \ref{conclusion}.

\begin{table*}
		\centering
		
		\caption[Gas density distribution properties]{\label{table:2} Electron distribution properties of the clusters analyzed by surface brightness.(2): The gas normalization factor. (3): The second gas normalization factor is applicable in a double beta profile. (4): The gas scale, where (5) gives the second gas scale for the double beta profile. (6) gives the exponent of the profile and the second exponent for the double beta profile is given in (7).  }
        \label{table:electrons} 
		\begin{tabular}{llllllll}
			\toprule
			Cluster name & $n_0\, (\mathrm{cm}^{-3})$ & $n_{0,\,2} \, (\mathrm{cm}^{-3})$ & $r_\mathrm{e} \,( \mathrm{kpc})$& $r_\mathrm{e,\,2} \, ( \mathrm{kpc})$& $\beta_\mathrm{e}$& $\beta_\mathrm{e,\,2}$& References \\
			(1)&(2)&(3)&(4)&(5)&(6)&(7)&(8)\\
			\hline
			Abell 4038 &0.022 &- &75.87 &- & -0.54&- & \cite{sun2007x} \cite{mohr1999properties}\\
			
			Abell 133 & & & & & & & \cite{morandi2014measuring} \\
			
			\hspace{2mm}Double beta profile &0.007&0.001&81.18&391&-0.75&-1.08&\\
			
			RXCJ0225.1-2928 & & & & & & &\cite{croston2008galaxy}\\
			\hspace{2mm}Beta profile &0.004&-&112&-&-0.8&&\\
		
			\hline
		\end{tabular}
	\end{table*}
 \section{Methodology}
 In order to determine upper limits for the DM cross-section the 2$\sigma$ exclusion is derived by comparing a predicted DM signal to the diffuse emission of a radio halo, or the value of the noise in the case of a non-detection. Therefore, it is required to remove the contributions of compact emission from the total measured flux. Contributions to the surface brightness from compact sources are identified with Python Blob Detector Source Finder (\texttt{PyBDSF}) \cite{mohan2015pybdsf} \footnote{https://github.com/lofar-astron/PyBDSF}, using the $7^{\prime\prime}$ primary beam corrected MGCLS data products. The superior resolution of these images allows \texttt{PyBDSF} to more accurately identify the boundaries of compact sources. The default island and boundary thresholds are utilized, $3 \sigma_{\mathrm{RMS}}$ and $5\sigma_{\mathrm{RMS}}$ respectively. The residual images are produced as an output of \texttt{PyBDSF}. 
 \subsection{Integrated flux comparison}
For galaxy clusters within the sample that contain a giant radio halo, we attempt to place an upper limit on the annihilation cross-section via a comparison of the measured integrated flux and the modelled DM flux over the region of the radio halo. A $2\sigma$ confidence level can be obtained through 
\begin{equation}
		\langle \sigma v \rangle = \frac{S+2\sigma_\mathrm{uncertainty}}{S^{\prime}_{\chi} \mathcal{B}} \; , 
	\end{equation}  
	where S is the measured flux density of the radio halo, $\sigma_\mathrm{uncertainty}$ is the uncertainty in the measured flux, $S^{\prime}_{\chi}$ is our modelled dark matter induced flux density as given in equation \ref{flux} where the cross section dependence has been extracted and $\mathcal{B}$ is the boost factor due to the presence of substructure within the dark matter halo (see equation \ref{boost}).

\begin{figure}
     \centering
     \includegraphics[width=0.97\linewidth]{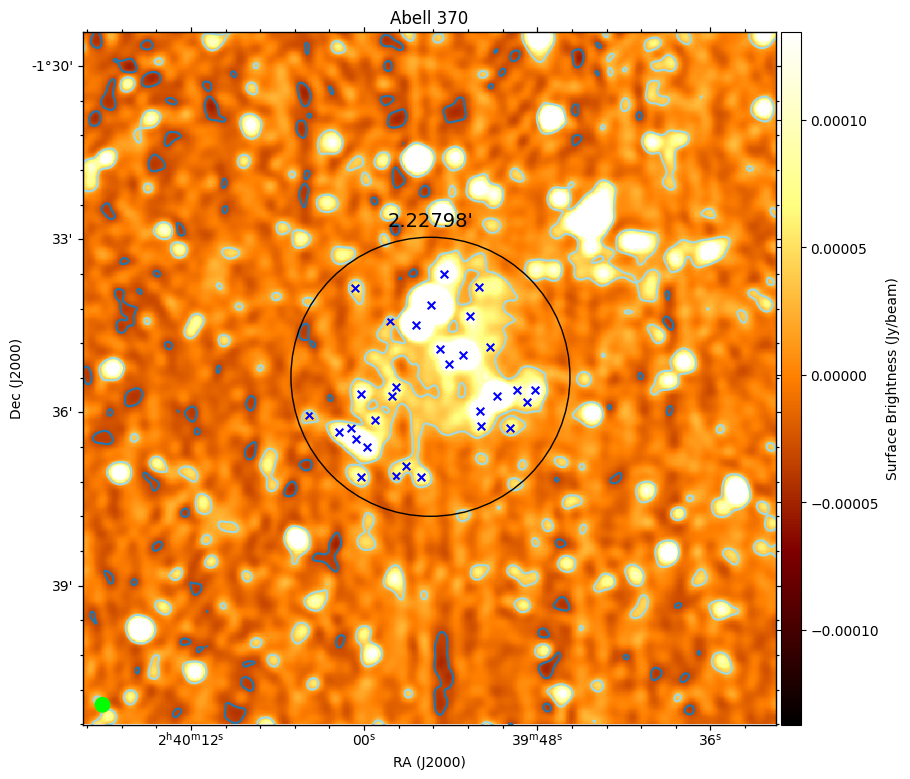}\\
     \includegraphics[width=0.97\linewidth]{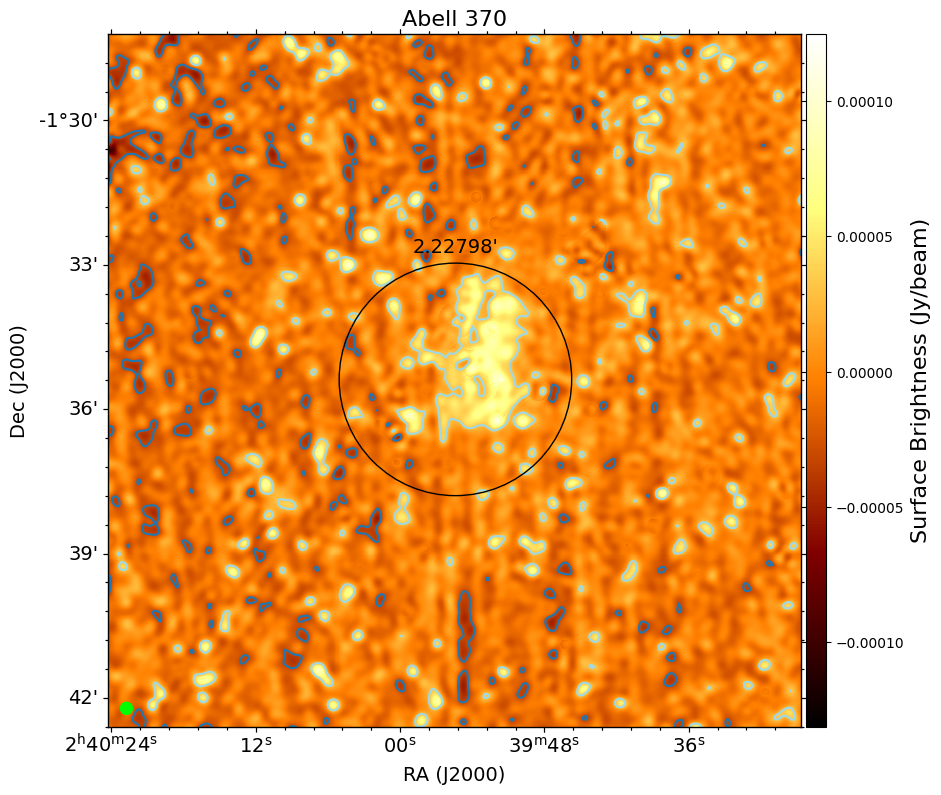}
     \caption{{Illustrative example of the procedure for obtaining the diffuse flux, depicting Abell 370. The contours shown are -3$\sigma$ and 3$\sigma$ in dark and light blue respectively, where $\sigma$ is the local RMS noise. The black circular region is the region chosen, that aims to contain the entire radio halo. The beam is shown in the lower left corner. \textit{Upper}: The surface brightness at $15^{\prime \prime}$ resolution. The crosses indicate the presence of point sources identified by \texttt{PYBDSF} in the full-resolution image, which are then removed. \textit{Lower}: The residual image produced by \texttt{PyBDSF} of region of the radio halo.}}
     \label{fig:A370_map}
 \end{figure}

 The region spanned by the radio halo is determined by applying a $3\sigma_{\mathrm{RMS}}$ contour, where $\sigma_{\mathrm{RMS}}$ is the local RMS noise. This is done on the $15^{\prime \prime}$ images, which are more sensitive to the faint diffuse emission. For simplicity, we choose our region of interest to be a circular region extended to contain the radio halo contours. An example of this is shown in Figure \ref{fig:A370_map}.  
 
 The integrated flux of the region of interest is obtained with the python plugin \texttt{radioflux} \footnote{https://www.extragalactic.info/\(\sim\)mjh/radio-flux.html}. The package can be used with the FITS image viewing software \texttt{SAOImageDS9} \footnote{https://sites.google.com/cfa.harvard.edu/saoimageds9} \cite{joye2003new} or stand-alone. This package calculates the total flux within a user-specified region by summing the values of all the pixels that fall within the region and applying a conversion factor of the area of the beam to a solid angle. The beam is assumed to be Gaussian, where its area is defined by the two-dimensional integral,
 \begin{equation}
		\Omega= \frac{\pi \, \theta_\mathrm{MAJ} \, \theta_\mathrm{MIN}}{4\ln 2} \; , 
	\end{equation}
	where $\theta_\mathrm{MAJ}$ and $ \theta_\mathrm{MIN}$ are the full-width half maximums (FWHMs) along the major and minor axes respectively, in terms of pixels \cite{condon2016essential}.
 
 Selecting the background subtraction option provided takes into account the noise of the image map, and estimates a statistical uncertainty.

 There are various sources of uncertainty that accumulate through this procedure, and it is non-trivial to accurately account for their effects. A statistical estimate of the uncertainty of the flux within the circular region is obtained by accounting for the noise. The systematic uncertainty is estimated to be $\sim 6\%$ of the measured flux, obtained via a flux density comparison of the MGCLS compact sources to the corresponding sources in the NRAO VLA Sky Survey (NVSS) and the Sydney University Molonglo Sky Survey (SUMSS) \cite{knowles2022meerkat}. A simple estimation of the uncertainty is obtained by adding these two quantities in quadrature, 
	\begin{equation}\label{eq:uncert}
		\sigma_\mathrm{uncertainty} = \sqrt{\sigma_\mathrm{statistical}^2 + \sigma_\mathrm{systematic}^2} \; . 
	\end{equation} 
    
An additional source of uncertainty comes from the identification of the compact sources by \texttt{PyBDSF}. We recall that \texttt{PyBDSF} attempts to fit Gaussian functions to bright islands of emission. However, it does not take into consideration that a peak in emission might have contributions from both compact sources and the diffuse emission present in the region. This can lead to over-subtraction when removing the contributions from compact sources. This uncertainty is non-trivial to quantify, but we make note that the effect of over-subtracting may be lowering the diffuse flux density measurement, and biasing our obtained upper limits on the annihilation cross-section to lower values.

 \subsection{Flux uncertainty exclusion limits}
For targets with a fainter diffuse background, we attempt a variation of a flux comparison. This approach requires modelling the DM annihilation signal and injecting this into the MGCLS map in the region of the cluster center. We note that the electron equilibrium distributions for this method are calculated using Green's method. This requires the assumption that the diffusion and loss functions have no spatial dependence, in contrast to a Crank-Nicolson scheme \cite{beck2022galaxy} which retains the spatial dependences of these functions. Beck et al \cite{beck2022galaxy} compare the results from the two methods and found that for integrated flux measurements, the assumption that the diffusion and loss functions do not have a spatial dependence does not overly affect the results. This is expected due to the large physical scales used to determine the integrated flux. 
In order to reduce the required computational time the MGCLS surface brightness plane is cropped to a $20^{\prime}$ by $20^{\prime}$ square centered on the pointing coordinates listed in Table \ref{table:1}. The contribution to the surface brightness from compact sources is removed with \texttt{PyBDSF} \footnote{https://pybdsf.readthedocs.io/en/latest/}. The signal injection is performed with the Common Astronomy Software Applications package \texttt{CASA} \cite{mcmullin2007casa}. Thereafter, we compare the flux within the region of interest before and after signal injection to determine the upper limits on the annihilation cross-section. However, we recall that dark matter halos, and therefore their signals, do not have a defined size. In order to choose an appropriate radius over which to measure the total flux we consider the relative size of the expected annihilation signal's dependency on the integration radius. We recall that the signal can be characterized by the J-factor, $J_\mathrm{ann}(\phi) = \int_{\mathrm{l.o.s}} \rho ^2(\phi,l)dl$, an integral of the square of the density distribution. Noting that the density distribution is spherically symmetric we can perform a volume integral of the profile to determine how much of the annihilation signal will be contained within a given radius. The contained signal is proportional to

	\begin{equation}
		\int_{0}^{a} \frac{x^2}{x^2(1+x)^4}= \frac{1}{3}- \frac{1}{3(a+1)^3} \; , 
	\end{equation} 
where $a =r/r_s$. From this, we are able to deduce that approximately 70\% of the anticipated dark matter signal is contained within $r=0.5r_s$, or $a=0.5$. We chose this to be the default choice for the region of integration.

The upper limit on the annihilation cross-section is then found through 
	\begin{equation}\label{eq:up}
		\langle \sigma v \rangle = 10^{-26} \mathrm{cm}^3 \mathrm{s}^{-1} \, \frac{2\sigma_\mathrm{uncertainty}}{S_\mathrm{Injected}-S_\mathrm{original}} \; , 
	\end{equation}
where $S_\mathrm{original}$ and $S_\mathrm{Injected}$ are the total fluxes in the same region before and after the injection of the DM signal, and $\sigma_\mathrm{uncertainty}$ is the statistical uncertainty of $S_\mathrm{original}$. Requiring that the excess flux after the injection is twice the uncertainty approximates a 95\% confidence level. The factor $10^{-26} $ assumes that the injection is performed with a modelled DM signal that has $\langle \sigma v \rangle =10^{-26}\ \mathrm{cm}^3\ \mathrm{s}^{-1}$.

The integrated flux is calculated by taking into account the noise via a background subtraction. For a significant DM signal in the background region, the required cross-section is approximately 2 orders of magnitude larger than the upper limits determined. Therefore ensuring that the background subtraction does not remove more than the intrinsic noise of the image. The result of this process is a set of exclusion curves for $\langle \sigma v \rangle$, shown in Figures \ref{fig:a133results}, \ref{fig:a4038results}, \ref{fig:rxcresults}, \ref{fig:rxcj0600results} and \ref{fig:rxcj0757results}.

 \subsection{Pixel-by-pixel method}

We also implement a statistical analysis of the MGCLS data that compares the radio surface brightness values at each individual pixel to the model values at corresponding sky locations. This method has been used in~\cite{sarkis2023radio}, and is similar to the one described in~\cite{regis2015local,regis2021emu}.

To compare these quantities at the pixel level, we first ensure that the MGCLS data and the calculated DM models are projected onto the same sky coordinates, using the \texttt{astropy} \texttt{reproject} module\footnote{https://reproject.readthedocs.io/en/stable/}. For each target, we select a square region of the sky centered on the MCXC central coordinates (as defined in Tab.~\ref{table:1}) and consider a Region of Interest (RoI) that is $2.5^{\prime} \times 2.5^{\prime}$ in angular size. After the removal of compact sources with the \texttt{PyBDSF} package, we mask any negative pixels that have an absolute value that is larger than 3 times the combined uncertainty estimate from the MGCLS plane cubes and the RMS output from \texttt{PyBDSF}, in a similar manner to~\cite{regis2021emu}. 

In each target, the RoI contains several thousand pixels, while the FWHM of the synthesised beam in each case contains $\sim 45$ pixels. If we consider the pixels inside a beamwidth to be highly correlated, while assuming those outside the beamwidth have no correlation, we are left with hundreds of effective resolution elements with which to do the analysis. Thus, under the afore-mentioned assumption, we adopt a Gaussian likelihood for our model, such that 
\begin{equation}\label{eqn:likelihood}
    \mathcal{L} = \mathrm{e}^{-\chi^2/2} \;,
\end{equation}
where 
\begin{equation}\label{eqn:chi2}
    \chi^2 = \dfrac{1}{N^{\mathrm{beam}}_{\mathrm{px}}}\sum_{n=1}^{N_{\mathrm{px}}}\left(\dfrac{S^{n}_{\mathrm{model}}-S^{n}_{\mathrm{data}}}{\sigma^{n}_{\mathrm{rms}}}\right)^2 .
\end{equation}

Here the standard $\chi^2$ statistic has been weighted by the number of pixels per beam ($N^{\mathrm{beam}}_{\mathrm{px}}$), which we use to approximately account for the correlation effects mentioned above, and the sum runs over all the pixels in the RoI ($N_{\mathrm{px}}$). In this case, the $\sigma_{\mathrm{rms}}^n$ values correspond to a combination of the MGCLS uncertainty and \texttt{PyBDSF} RMS estimates. Our free parameter in this analysis is the thermal annihilation relic cross-section $\langle \sigma v \rangle$, which gives us a set of $S_{\mathrm{model}}$ values with which to calculate $\chi^2$. We then find a one-sided upper confidence level for $\langle \sigma v \rangle$ by performing a standard likelihood ratio test (LRT) over each model (denoted by the subscript $i$):
\begin{align}\label{eqn:lrt}
    \lambda_c &= -2\ln\left[\mathcal{L}_i / \mathcal{L}_{\mathrm{max}}\right] \;,\nonumber \\
    \Rightarrow \lambda_c &= \chi^2_i - \chi^2_{\mathrm{min}}\;.
\end{align}
Since the quantity $\chi^2_{\mathrm{min}}$ here is the one which maximizes the likelihood, we can manipulate the value of $\lambda_c$ to define the extent of the confidence level; for example, a value of $\lambda_c = 0$ will result in a best-fit model. In this work, we consider a $2\sigma$ confidence level, which means that in the probability function that describes the likelihood, 
\begin{equation}\label{eqn:probability}
    P = \int_{\sqrt{\lambda_c}}^{\infty} \; \mathrm{d}\chi \frac{\exp(-\chi^2/2)}{\sqrt{2\pi}} = 0.05 \;,
\end{equation}
which yields $\lambda_c = 2.71$. We use this in Eqn~\ref{eqn:lrt} to find an upper-limit model for $\langle \sigma v \rangle$, which we then repeat for each WIMP mass and DM parameter set. The result of this process is a set of exclusion curves for $\langle \sigma v \rangle$, shown in Figures \ref{fig:a133results}, \ref{fig:a4038results}, \ref{fig:rxcresults}, \ref{fig:rxcj0600results} and \ref{fig:rxcj0757results}. 

The statistical tests employed are valid for the amount of data points used in each image. In addition, the analysis is performed in the same way for all the images. By utilizing every available pixel within the RoI and comparing to the model in a direct way, the analysis can be considered robust. It is important to note that, unlike \cite{regis2015local,regis2021emu}, we do not profile the likelihood. Instead, we consider a range of upper limits resulting from the choice of halo profile configurations and separately consider the effect of the magnetic field strength.  
 
\section{Results}\label{section:Results}

In this investigation, WIMP masses are probed from 10 GeV to 1000 GeV. For the 2HDM+S model the lowest mass that can be probed is 75 GeV for the annihilation chain studied, see section \ref{section:2hdm+s}. This is the lowest DM mass that is able to produce the mediator boson, or the heavy scalar boson together with a Higgs boson. The effect of this is evident by the reduced range of the 2HDM+S cross-section limits in the results.  

\subsection{Integrated flux comparison}
The upper limits of the annihilation cross-section for clusters containing giant radio halos assume an NFW DM density profile. In addition, the electron distributions are obtained through the use of Green's functions.   

\begin{figure}
    \centering
    \includegraphics[width=\linewidth]{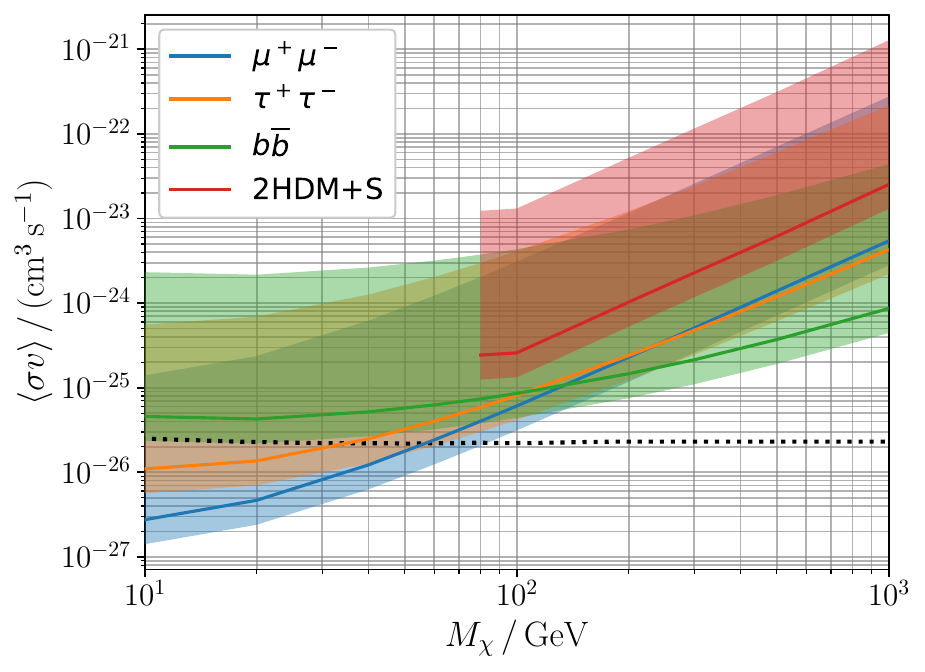}
    \caption{Upper limits on the DM annihilation cross section at a 95\% confidence level for Abell 370, determined the integrated flux comparison of the DM signal and the diffuse emission of the radio halo. An NFW profile is assumed to produce the median case limits (shown by the solid lines), while the lower bound of the uncertainty band accounts for the upper limits of the scale radius and density, while the upper bound of the band utilises the lower bound of $r_s$ and $\rho_s$, a modified NFW density profile ($\alpha$=0.5), as well as reducing $B_0$  from 5 to 2$\mu$G. The dotted line represents the calculated thermal relic value of the cross-section, taken from \cite{steigman2012precise}., below which models are excluded.}
    \label{fig:A370results}
\end{figure}
The results for Abell 370 are depicted in Figure \ref{fig:A370results}. These are the most constraining results of the twelve clusters investigated. With the median case NFW limits we are able to exclude annihilation through $\mu^+\mu^-$ below 60 GeV, and through $\tau^+\tau^-$ below 30 GeV.

More constraining limits on DM properties can be obtained from clusters with giant radio halos by removing the contribution to the flux density due to normal diffuse cosmic rays. This would require radio continuum spectral data and has been attempted by  Chan et al \cite{chan2020constraining,chan2019fitting, chan2020possible}. In these works, it is argued that radio continuum spectrum data could be used to differentiate the contributions of cosmic rays and dark matter to the diffuse emission as the spectral index could be different. However, the spectral shape of the cosmic ray contribution depends on the assumed model for the production mechanism, primary electron emission models, secondary electron emission models, or in situ acceleration. Parametric equations \cite{chan2020constraining} for the cosmic ray contributions  are:
\begin{equation}
	S_\mathrm{CR}= S_\mathrm{CR, \,0} \left(\frac{\nu}{\mathrm{GHz}} \right)^{-\alpha} \left[\frac{1}{1+\left(\frac{\nu}{\nu_s} \right)^{\Gamma}} \right], \, \Gamma= 0.5 \, \mathrm{or}\,  1 \; ,
\end{equation} 
as well as
\begin{equation}
	S_\mathrm{CR}= S_\mathrm{CR, \,0} \left(\frac{\nu}{\mathrm{GHz}} \right)^{-\alpha} \; ,
\end{equation}
and
\begin{equation}
	S_\mathrm{CR}= S_\mathrm{CR, \,0} \left(\frac{\nu}{\mathrm{GHz}} \right)^{-\alpha} \exp^{(-\nu ^{1/2}/v_s^{1/2})} \; ,
\end{equation}
respectively. In the above $S_\mathrm{CR, \,0}$, $ \nu_s$ and $\alpha$ are free parameters for the fitting to the observed radio spectrum. Such an analysis assumes that the measured flux with its uncertainty is the expected flux density. This is valid in cases where the contribution from star formation is negligible \cite{heesen2021comment} but should be used with caution, particularly for smaller-scaled structures where the contribution from star formation is significant.

\subsection{Comparison of flux uncertainty and pixel-by-pixel method}
For the clusters investigated through these methods, we have depicted our median limits, where the median values of the dark matter halo parameters are utilised, with a standard NFW profile, and the default magnetic field strength of either 5 $\mu$G or 8$\mu$G for non-cool-cored and cool-cored clusters respectively. The lower bound of the uncertainty band is found with the upper bound on the scale density while maintaining the default magnetic field strength and an NFW profile. The upper bound of the annihilation cross-section limits is obtained with the lower bound on the scale density, a modified NFW profile ($\alpha=0.5$), and a reduced magnetic field strength of 2$\mu$G. We assume that for any reasonable modelling parameters, the upper limits on the annihilation cross-section will lie within the band of uncertainty. We note that the shallower dark matter density profile weakens the upper limits on the annihilation cross-section by a factor of 2, while the reduction in $B_0$ weakens the limits by a factor 7 (10 for cool-cored) or 4 for the flux uncertainty method and the pixel-by-pixel method respectively. Therefore it can be seen that the chosen central magnetic field value can induce a greater uncertainty on the derived upper limits than the uncertainty from the shape of the dark matter density profile.  

We display the results for both flux uncertainty and pixel-by-pixel methods for Abell 133, Abell 4038, RXCJ0225.1-2928, J0600.8-5835, and J0757.7-5315. The bounds on the annihilation cross section obtained through the flux uncertainty method solve for the equilibrium electron distributions through the use of Green's functions. We do not expect the bounds on the annihilation cross-section to vary significantly if a Crank-Nicolson scheme is used instead when utilizing an integrated flux value. The bounds on the annihilation cross-section obtained through the pixel-by-pixel method make use of the OS method to solve for the electron distributions.

\begin{figure}
    \centering
    \includegraphics[width=\linewidth]{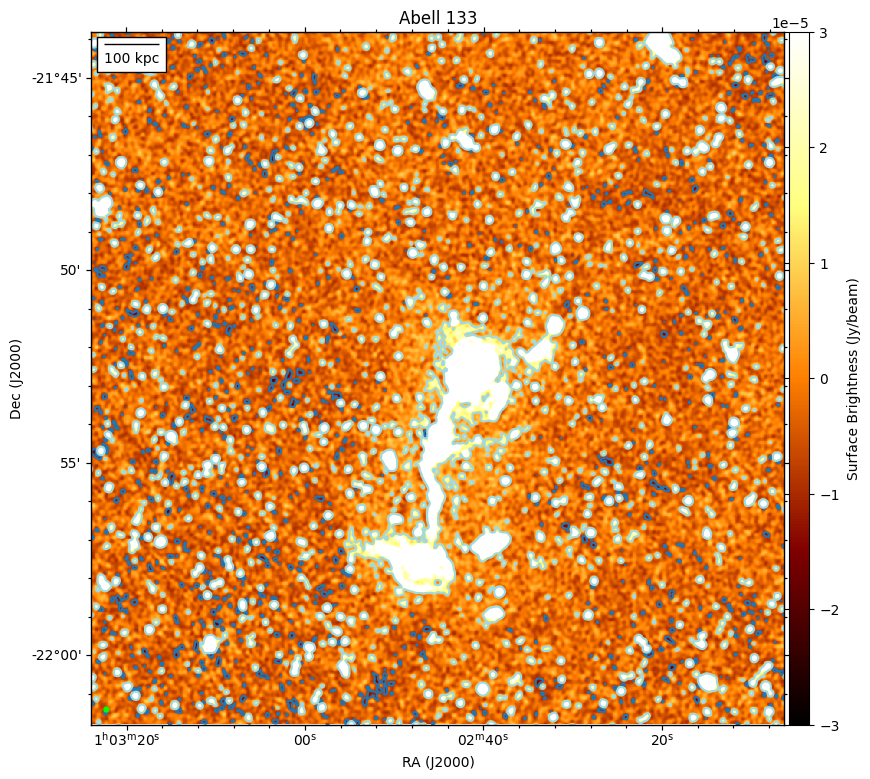}\\
    \includegraphics[width=\linewidth]{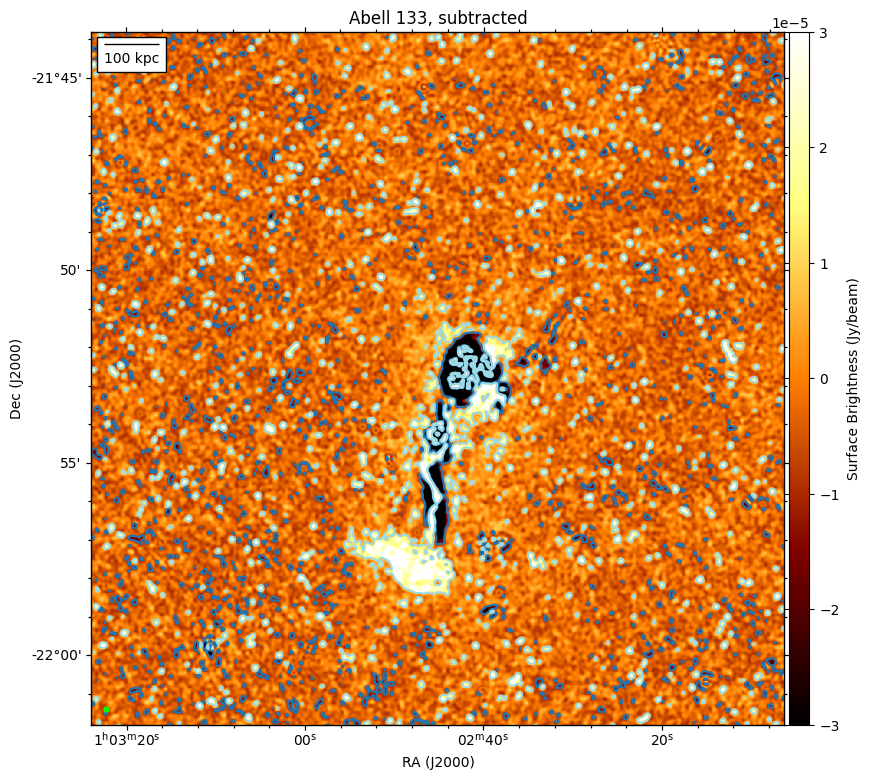}
    \caption{Abell 133 focused on the MeerKAT pointing coordinates, cropped to a $20^{\prime} \times 20^{\prime}$ square. The beam size is shown in the lower left corner of all maps. The dark and light contours correspond to $-2\sigma$ and $2\sigma$ of the local average RMS respectively.  \textit{Upper}: The MGCLS $7^{\prime \prime}$ surface brightness map. \textit{Lower}: The residual image produced by \texttt{PyBDSF} showing  over-subtraction.} 
    \label{fig:a133}
\end{figure}

\begin{figure}
    \centering
    \includegraphics[width=\linewidth]{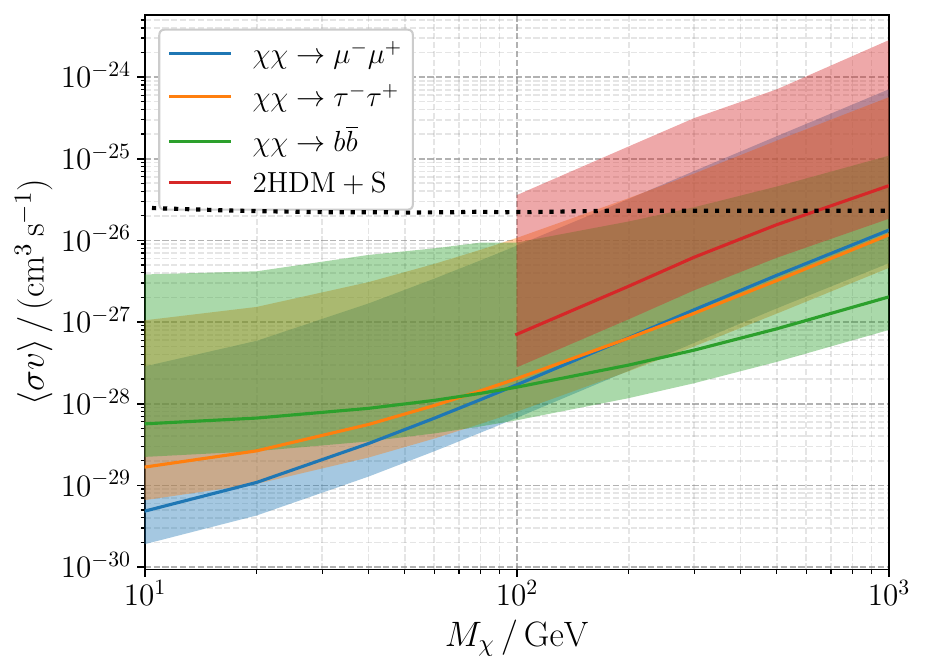}\\
    \includegraphics[width=\linewidth]{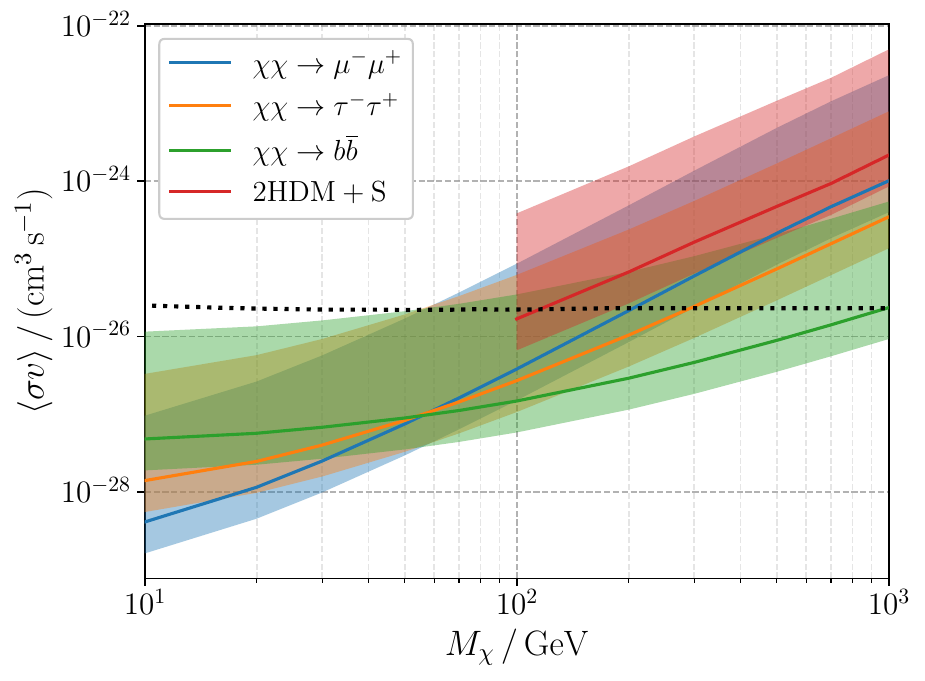}
    \caption{Upper limits ($2\sigma$) on the annihilation cross-section for WIMPs in Abell 133, annihilating via 3 generic intermediate channels as well as through 2HDM+S. The solid line represents the limits obtained with the median value of the scale dark matter parameters, with the default $B_0$ value ($8\mu$G) and an NFW density profile. The uncertainty bands consider the uncertainty in the calculated dark matter scale parameters, the uncertainty in the slope of the density profile, and the uncertainty in the magnetic field. The upper bound of the uncertainty band is found with the lower values of the scale density, a modified NFW profile with $\alpha=0.5$, and a reduced magnetic field strength $B_0=2 \mu$G, while the lower bound considers the default magnetic field, an NFW profile and the upper value of the scale density. The dotted line represents the calculated thermal relic value, taken from \cite{steigman2012precise}. \textit{Upper}: Results obtained through the flux uncertainty comparison. \textit{Lower}: Results obtained through the pixel-by-pixel method.  }
    \label{fig:a133results}
\end{figure}

\begin{figure}
    \centering
    \includegraphics[width=\linewidth]{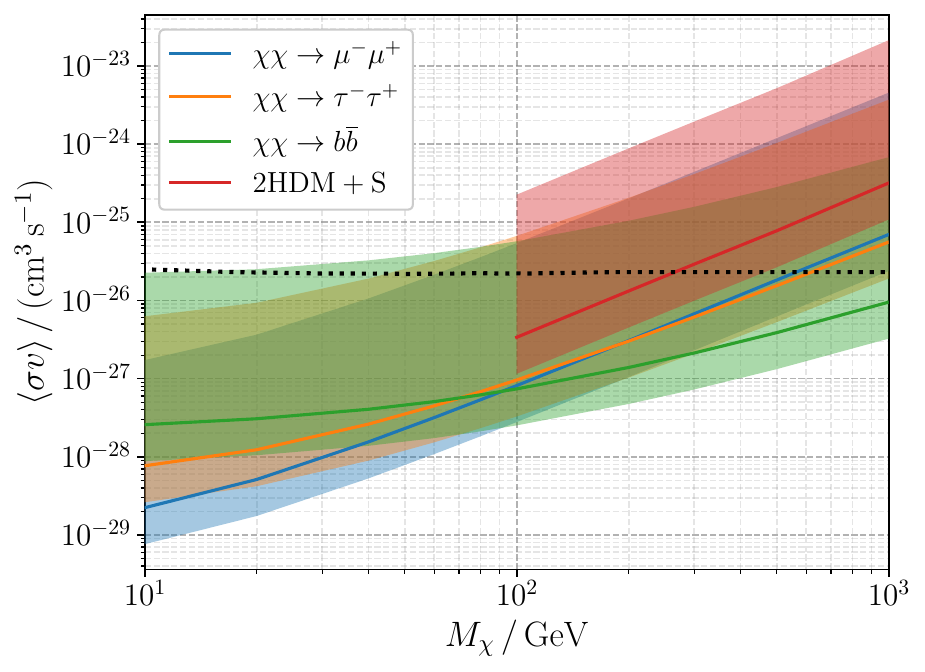}\\
    \includegraphics[width=\linewidth]{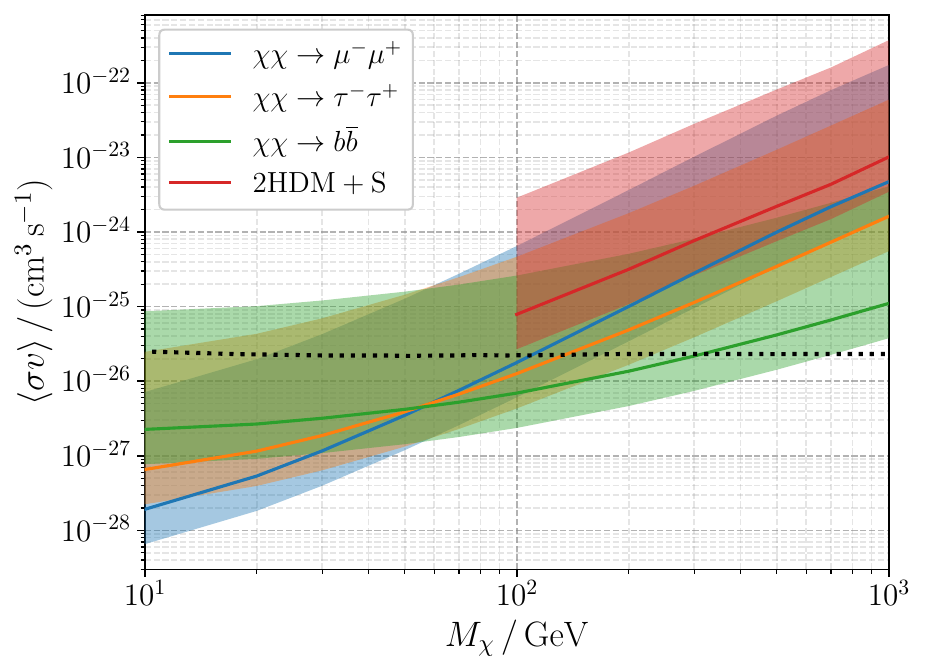}
    \caption{Upper limits ($2\sigma$) on the annihilation cross-section for WIMPs in Abell 4038, annihilating via 3 generic intermediate channels as well as through 2HDM+S. The solid line represents the limits obtained with the median value of the scale dark matter parameters, with the default $B_0$ value ($8\mu$G) and an NFW density profile. The uncertainty bands consider the uncertainty in the calculated dark matter scale parameters, the uncertainty in the slope of the density profile, and the uncertainty in the magnetic field. The upper bound of the uncertainty band is found with the lower values of the scale density, a modified NFW profile with $\alpha=0.5$, and a reduced magnetic field strength $B_0=2 \mu$G, while the lower bound considers the default magnetic field, an NFW profile and the upper value of the scale density. The dotted line represents the calculated thermal relic value, taken from \cite{steigman2012precise}. \textit{Upper}: Results obtained through the flux uncertainty comparison. \textit{Lower}: Results obtained through the pixel-by-pixel method.}
    \label{fig:a4038results}
\end{figure}

\begin{figure}
    \centering
    \includegraphics[width=\linewidth]{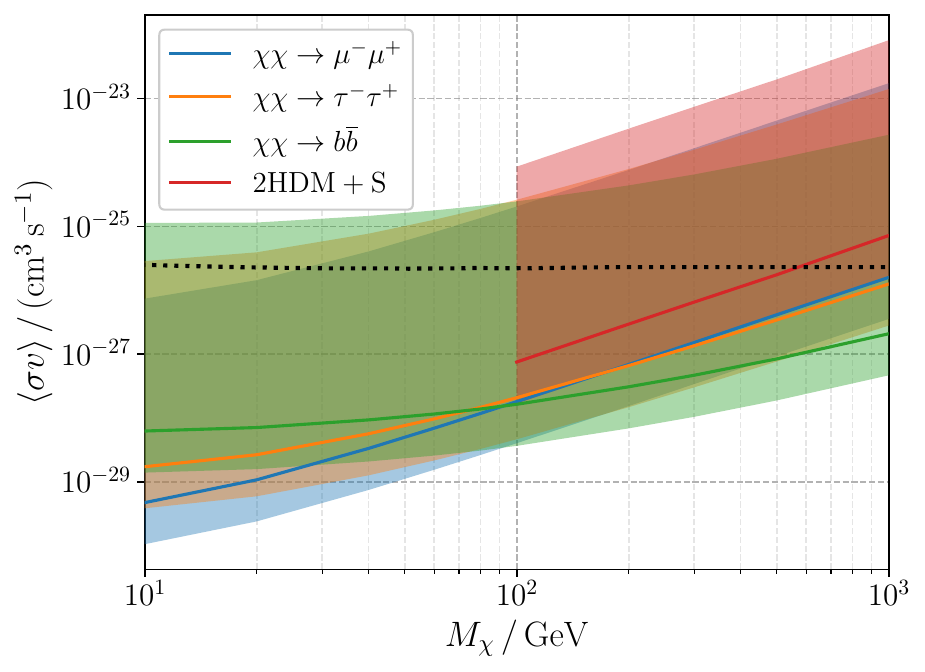}\\
    \includegraphics[width=\linewidth]{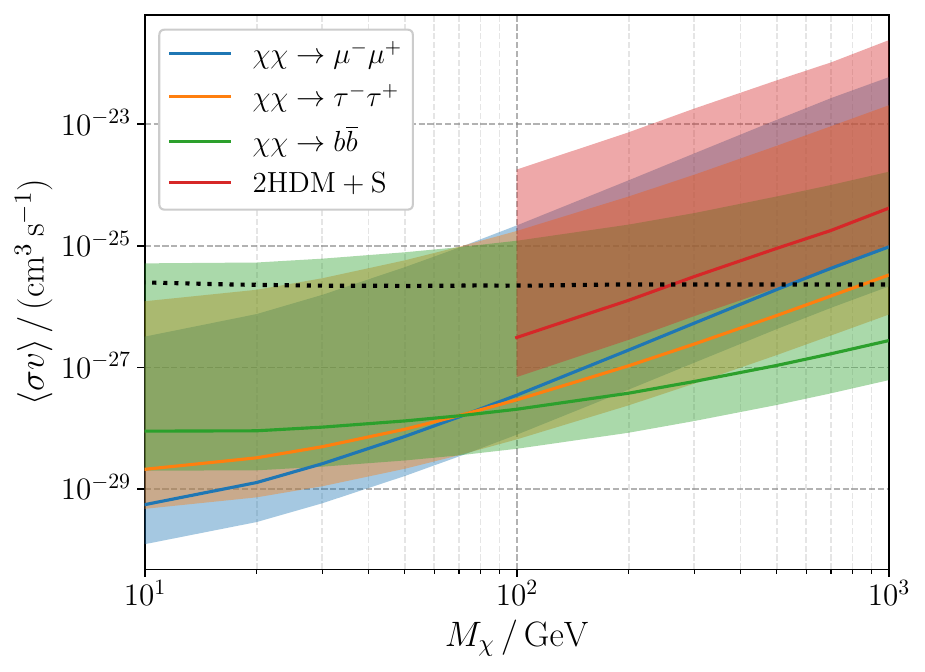}
    \caption{Upper limits ($2\sigma$) on the annihilation cross-section for WIMPs in RXCJ0225.1-2928, annihilating via 3 generic intermediate channels as well as through 2HDM+S. The solid line represents the limits obtained with the median value of the scale dark matter parameters, with the default $B_0$ value ($5\mu$G) and an NFW density profile. The uncertainty bands consider the uncertainty in the calculated dark matter scale parameters, the uncertainty in the slope of the density profile, and the uncertainty in the magnetic field. The upper bound of the uncertainty band is found with the lower values of the scale density, a modified NFW profile with $\alpha=0.5$, and a reduced magnetic field strength $B_0=2 \mu$G, while the lower bound considers the default magnetic field, an NFW profile and the upper value of the scale density. The dotted line represents the calculated thermal relic value, taken from \cite{steigman2012precise}. \textit{Upper}: Results obtained through the flux uncertainty comparison. \textit{Lower}: Results obtained through the pixel-by-pixel method.}
    \label{fig:rxcresults}
\end{figure}

\begin{figure}
    \centering
    \includegraphics[width=\linewidth]{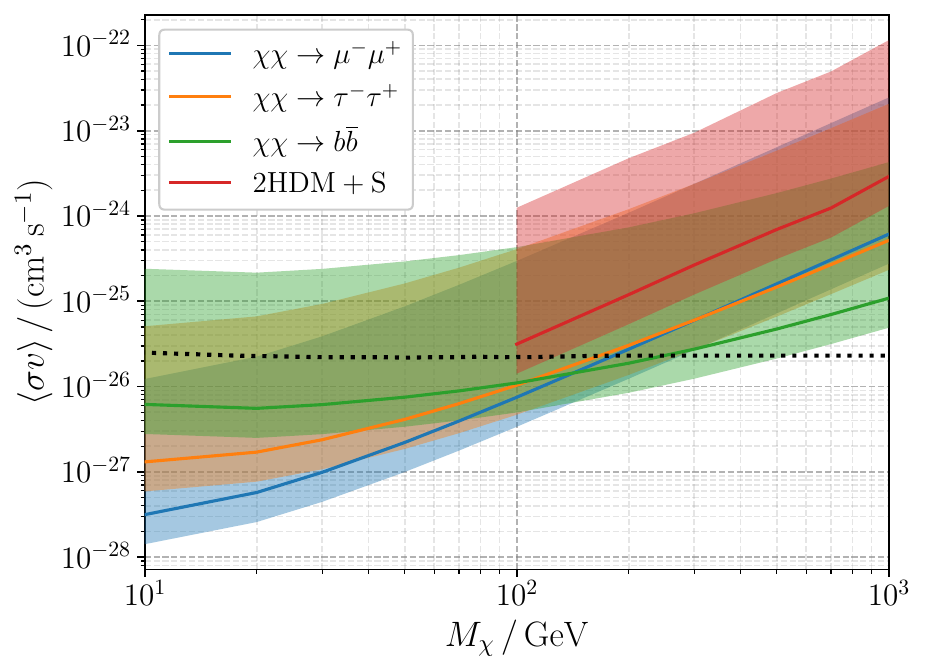}\\
    \includegraphics[width=\linewidth]{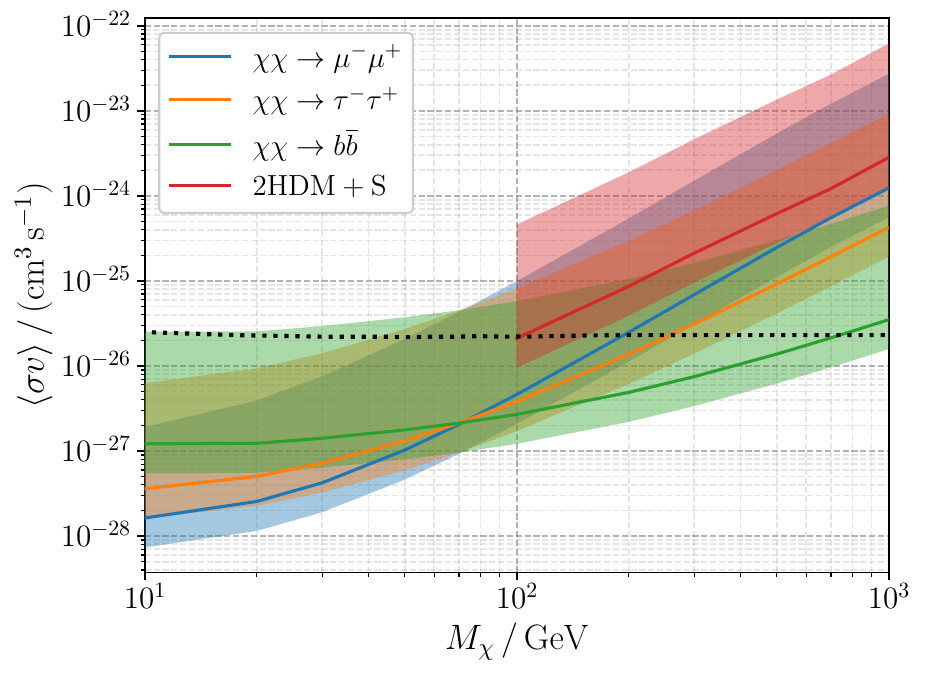}
    \caption{Upper limits ($2\sigma$) on the annihilation cross-section for WIMPs in J0600.8-5835, annihilating via 3 generic intermediate channels as well as through 2HDM+S. The solid line represents the limits obtained with the median value of the scale dark matter parameters, with the default $B_0$ value ($5\mu$G) and an NFW density profile. The uncertainty bands consider the uncertainty in the calculated dark matter scale parameters, the uncertainty in the slope of the density profile, and the uncertainty in the magnetic field. The upper bound of the uncertainty band is found with the lower values of the scale density, a modified NFW profile with $\alpha=0.5$, and a reduced magnetic field strength $B_0=2 \mu$G, while the lower bound considers the default magnetic field, an NFW profile and the upper value of the scale density. The dotted line represents the calculated thermal relic value, taken from \cite{steigman2012precise}. \textit{Upper}: Results obtained through the flux uncertainty comparison. \textit{Lower}: Results obtained through the pixel-by-pixel method.}
    \label{fig:rxcj0600results}
\end{figure}

\begin{figure}
    \centering
    \includegraphics[width=\linewidth]{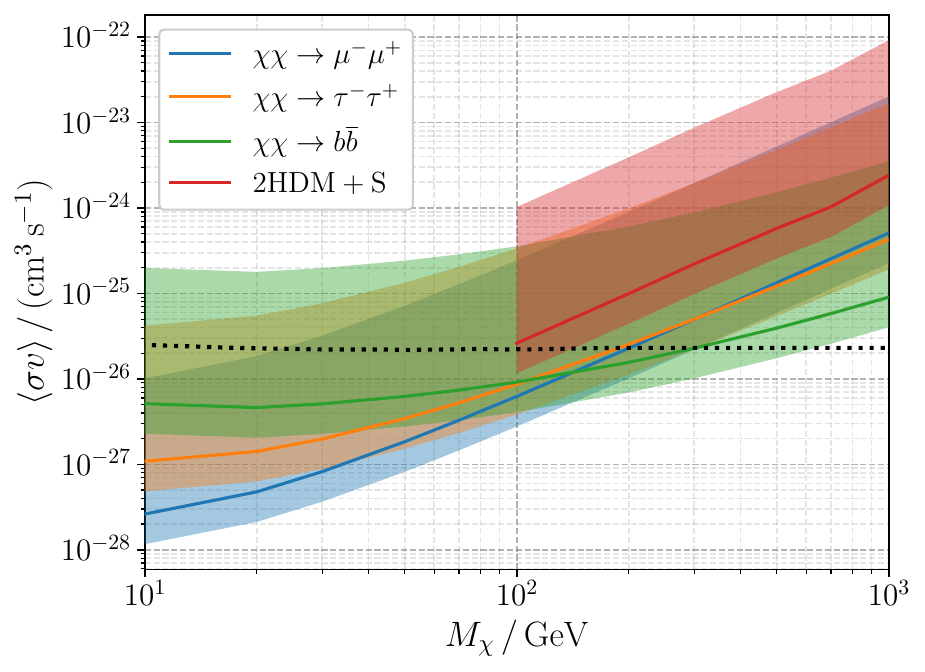}\\
    \includegraphics[width=\linewidth]{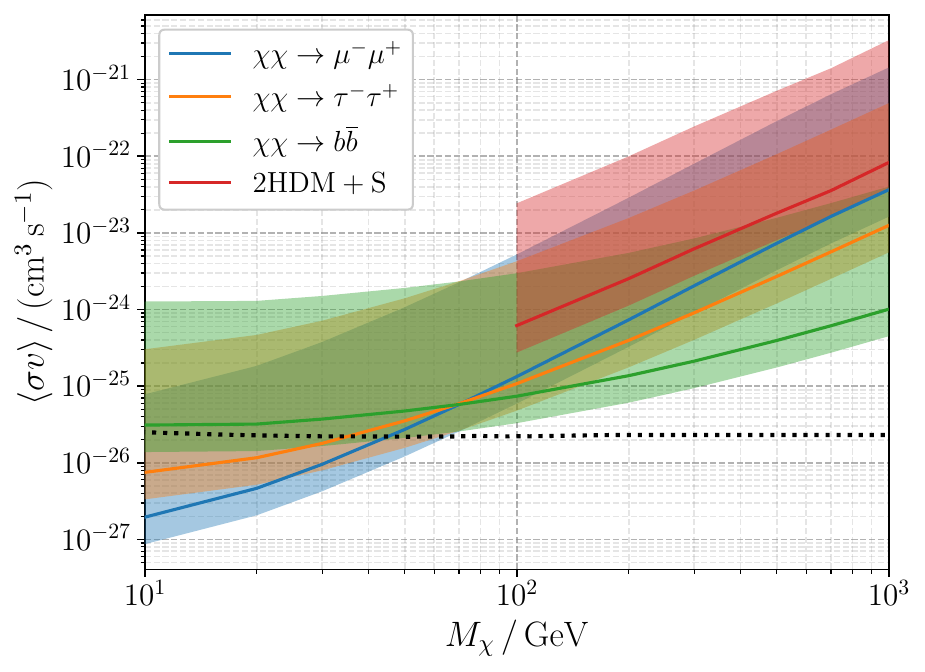}
    \caption{Upper limits ($2\sigma$) on the annihilation cross-section for WIMPs in J0757.7-5315, annihilating via 3 generic intermediate channels as well as through 2HDM+S. The solid line represents the limits obtained with the median value of the scale dark matter parameters, with the default $B_0$ value ($5\mu$G) and an NFW density profile. The uncertainty bands consider the uncertainty in the calculated dark matter scale parameters, the uncertainty in the slope of the density profile, and the uncertainty in the magnetic field. The upper bound of the uncertainty band is found with the lower values of the scale density, a modified NFW profile with $\alpha=0.5$, and a reduced magnetic field strength $B_0=2 \mu$G, while the lower bound considers the default magnetic field, an NFW profile and the upper value of the scale density. The dotted line represents the calculated thermal relic value, taken from \cite{steigman2012precise}. \textit{Upper}: Results obtained through the flux uncertainty comparison. \textit{Lower}: Results obtained through the pixel-by-pixel method.}
    \label{fig:rxcj0757results}
\end{figure}

For Abell 4038, RXCJ0225.1-2928, J0600.8-5835, and J0757.7-5315 the residual image produced by \texttt{PyBDSF} is used as the sky image, either as the initial image for signal injection or pixel-by-pixel comparison. The central region of Abell 133 contains an extremely bright radio galaxy, with an extended radio tail. When this contribution was removed with \texttt{PyBDSF} this resulted in extreme over-subtraction, see Figure \ref{fig:a133}. As a consequence, this produced nonphysical values for the surface brightness and the integrated flux. For this reason, we keep the contribution of this radio galaxy intact for the flux uncertainty method. As significantly negative pixels are masked in the pixel-by-pixel method, this procedure makes use of the residual image. The results for Abell 133, Abell 4038, RXCJ0225.1-2928, J0600.8-5835 and J0757.7-5315 are depicted in Figures \ref{fig:a133results}, \ref{fig:a4038results}, \ref{fig:rxcresults},\ref{fig:rxcj0600results} and \ref{fig:rxcj0757results}.   

\begin{figure}[t!]
    \centering
    \includegraphics[width=\linewidth]{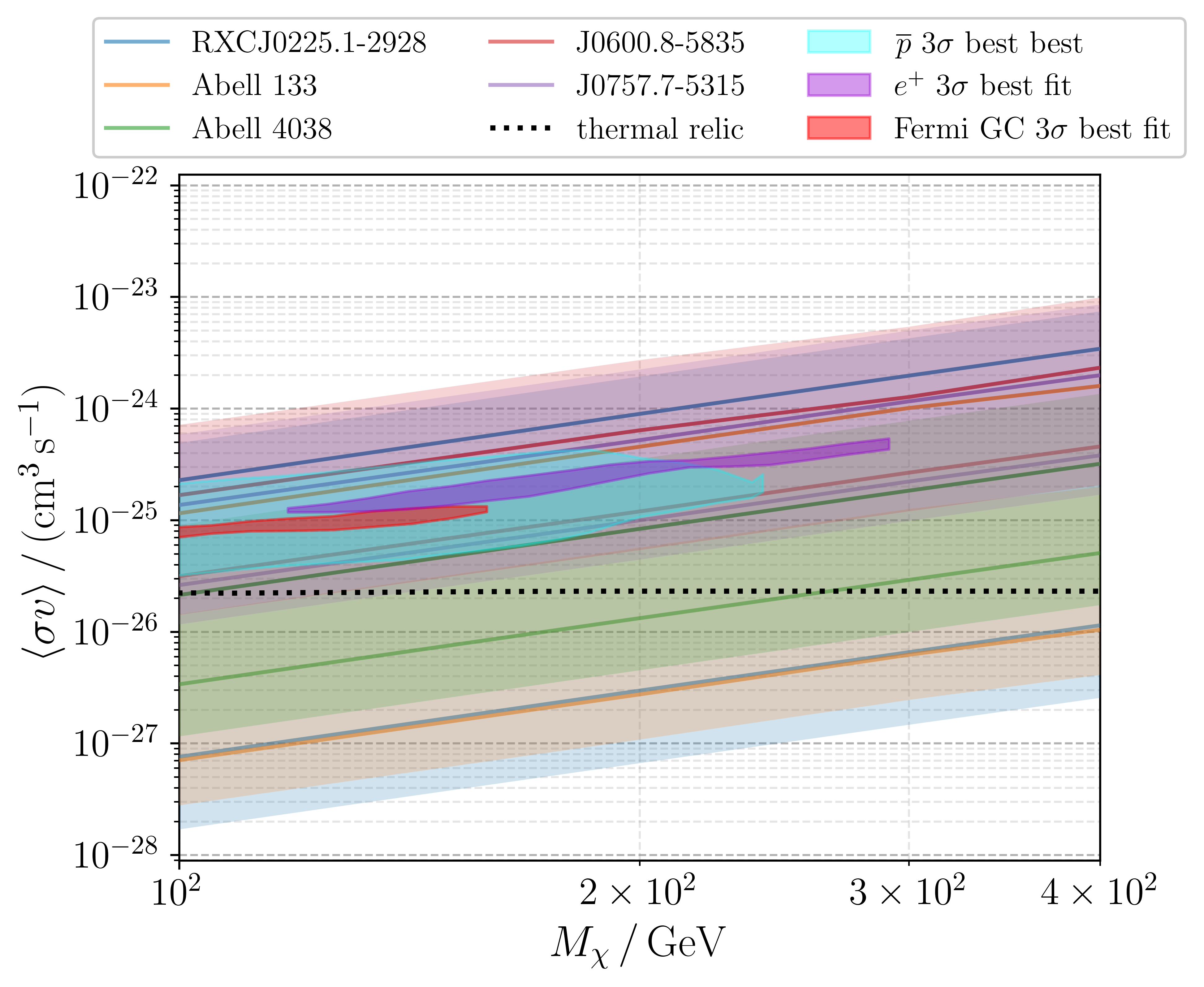}\\
    \includegraphics[width=\linewidth]{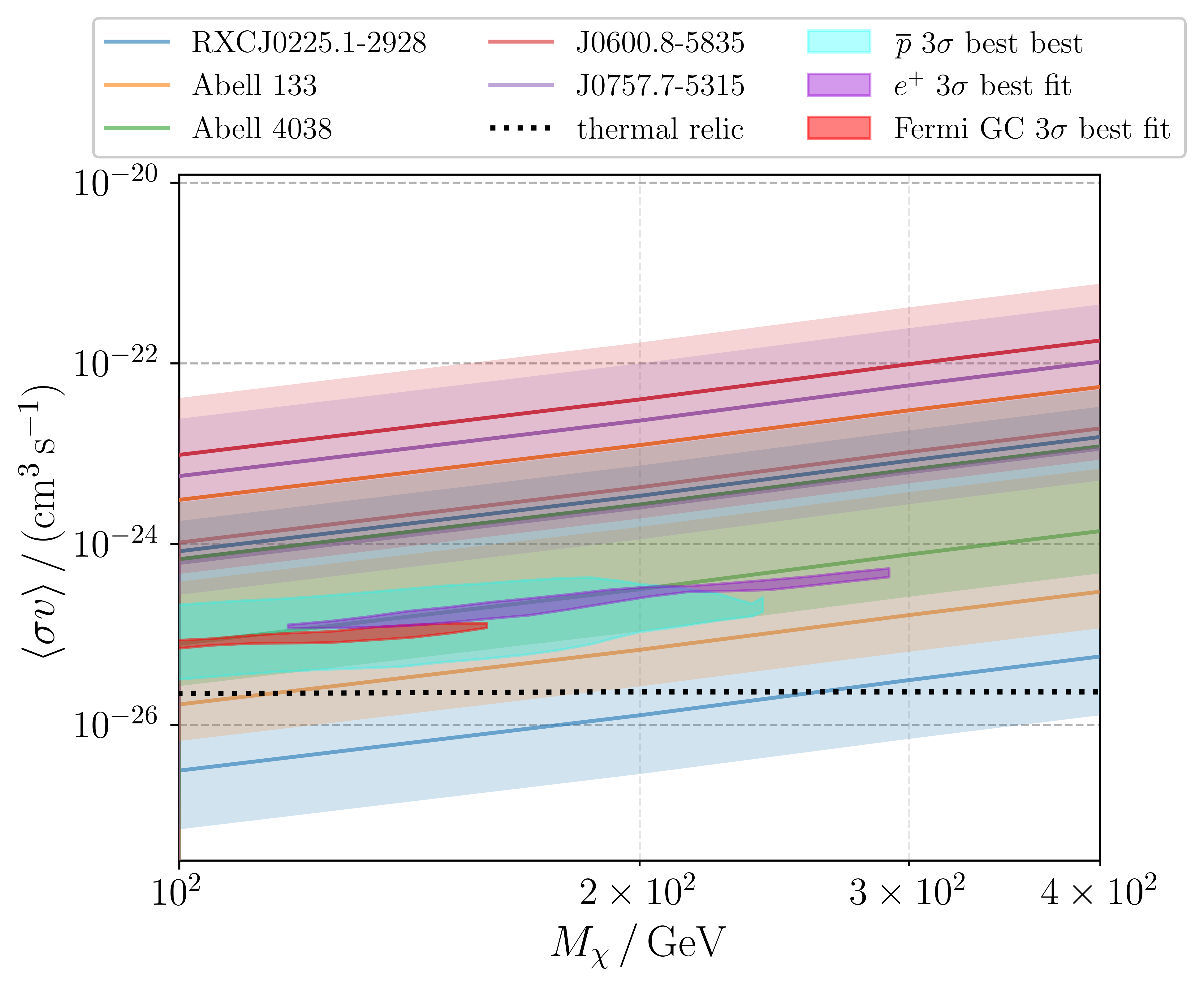}
    \caption{Upper limits of the dark matter annihilation cross section determined at a 95\% confidence level for the 2HDM+S annihilation channel overlaid with the parameter space fittings for the positron, anti-proton, and gamma-ray excesses determined by Beck et al \cite{beck2021connecting}. The solid coloured lines indicate the upper limits in the median case. The dotted line represents the thermal relic value, below which models can be excluded. \textit{Upper}: Upper limits obtained through the flux uncertainty method. \textit{Lower}: Upper limits obtained through the pixel-by-pixel method. }
    \label{fig:2hdms}
\end{figure}

The DM candidate within the 2HDM+S particle physics model is of interest as the mass range of this candidate overlaps with that of DM models for various astrophysical excesses. At present, the only constraints on this candidate have been determined by a parameter space fitting of $\chi \overline{\chi} \rightarrow \, S \, \rightarrow X$
to the AMS-02 positron data performed by Beck et al \cite{beck2021connecting}, and the overlapping regions of the anti-proton and the Fermi-LAT galactic center gamma-ray excess parameter spaces. In Figure \ref{fig:2hdms} we display these fittings with the annihilation cross-section constraints for the five galaxy clusters investigated. The results show significant overlap for both the flux uncertainty method and the pixel-by-pixel method. As such, the DM candidate within the 2HDM+S model remains a viable explanation for the astrophysical excesses, especially considering the uncertainties of the modelling parameters. The results show that with MeerKAT we are able to begin to probe the 2HDM+S parameter space.  These results exceed the relatively crude sensitivity predictions made for 100 hours of MeerKAT observations for Reticulum II~\cite{beck2021connecting}. This is achieved with less than 10 hours of on-target observation with actual MeerKAT data.

\section{Discussion and Conclusion}\label{conclusion}
The constraints produced in this analysis of three galaxy clusters (A4038, A133, and RXCJ0225.1-2928) using a cuspy NFW DM halo profile, with median values of the modelling parameters, are comparable to the constraints produced by Regis et al \cite{regis2021emu} for the Large Magellanic Cloud (LMC). Results within the uncertainty bands are comparable to those found in the literature for various astrophysical targets  \cite{egorov2022updated,chan2019ruling,basu2021stringent}. The most constraining results are produced for Abell 4038 and RXCJ0225.1-2928 thereafter. We find that, even within the uncertainties, the upper limits for Abell 133 are superior to those determined with Fermi-LAT data of a sample of 49 clusters \cite{ando2012fermi}. For the less massive clusters, J0600.8-5835 and J0757.7-5315, the median case limits are superior to those determined with gamma-ray data obtained with Fermi-LAT \cite{ando2012fermi,hoof2020global,gammaldi2021dark,calore2018dark}, while remaining comparable within the uncertainties. 
Notably, the pixel-by-pixel method produces weaker results than the integrated flux case. This should be attributed to the background subtraction performed for the integrated analysis. Importantly, we have confirmed that this process did not over-subtract via the inclusion of significant DM flux. In further work, we will attempt to improve the pixel-by-pixel analysis to provide additional robust confirmation of the integrated flux results. Additionally, the results from \cite{regis2021emu} are subject to some recent uncertainty, as the Milky-Way mass, and thus that of the LMC too, may be three to five times smaller~\cite{Jiao:2023aci} than the estimate used in \cite{regis2021emu}. This would very probably weaken the limits from \cite{regis2021emu} by an order of magnitude.

In galaxy clusters with extended source emission with a complicated morphology, \texttt{PyBDSF} with the standard parameters is unable to remove its contribution to the surface brightness adequately. This can result in an unrealistic number of pixels with a significantly negative flux value. Future studies may therefore benefit from a compact source subtraction procedure that is tailored to complex morphologies.

An analysis of galaxy clusters is able to overcome some of the uncertainties faced by studies on more common targets, such as dwarf Spheroidal galaxies (dSphs). In galaxy clusters, the uncertainties of the modeling parameters are significantly reduced, in particular the magnetic field structure and the effects of diffusion. Consider the effects of the chosen diffusion constant. In dwarf spheroidal galaxies, very little is known about the value of $D_0$ for dwarf spheroidal galaxies, due to their low luminosity. It can be seen that varying the diffusion coefficient and magnetic field strength over 2 orders of magnitude causes the derived upper limits on the cross-section to vary two to three orders of magnitude, for example ~\cite{gajovic2023weakly}. In contrast, the upper limits on the cross-section in galaxy clusters are more robust to variations in this constant where the difference in the derived upper limits is negligible \cite{beck2022galaxy,sarkis2023radio}. When considering the magnetic field it is evident that the uncertainties are greatly reduced in galaxy clusters. The magnetic field strength in dwarf spheroidal galaxies can be varied over as much as three orders of magnitude ~\cite{gajovic2023weakly,kar2020heavy} as it is extremely difficult to gain observational insight through polarization studies, and instead the magnetic field strengths are typically estimated through an empirical scaling relation with the star formation rate ~\cite{regis2017dark}. In contrast, recent studies place cluster magnetic field strengths within a factor of 5 ~\cite{govoni2004magnetic,osinga2022detection}.

The uncertainties can be further reduced by a tailored observation of a source for which the multi-wavelength archival data is available, with longer exposure times. However, one key uncertainty faced is the strength of magnetic fields within clusters.

Faraday rotation measures are one of the most powerful methods of constraining magnetic fields. The observation of background or embedded sources of galaxy clusters will be able to probe the magnetic field properties. MeerKAT has demonstrated its ability to sample large sources at a high resolution and derive polarization and spectral information with only a few hours of observational data \cite{de2022meerkat}. These capabilities will become more powerful with the SKA \cite{bonafede2015unravelling}, allowing for detailed magnetic field constraints for many galaxy clusters. This will allow for a more accurate analysis, and potentially strengthen the constraints on the DM parameter space.  This advantage may not be available in dwarf galaxies, due to their low luminosity and small size, any polarized emission may be too weak to be detected \cite{beck2012magnetic}.

In this work, we have injected the DM signal directly into the image plane of the clusters considered. Typically modeled signals are injected into the visibility plane. The linear nature of the Fourier transform between the planes means that it should be possible to perform the injection in either the image or the visibility plane. However, due to the nature of interferometers, the transform is not ideal. If the injection is performed in the visibility space, the recovered flux of an injected source depends largely on the \textit{uv} sampling. As any measured visibility function is a limited subset of the true visibility function it is unlikely to recover the full signal from an injected source. Thus, if the signal was injected into the visibility plane we expect to recover a weaker signal in the image plane, and therefore obtain less constraining upper limits of the annihilation cross section. This effect will be compensated for by performing the compact source subtraction in the visibility plane, which will reduce the effects of over-subtraction. In the upcoming SKA era, visibilities are unlikely to be available, due to the high volumes of data collected. As such, improvements to image plane source subtraction are required for future studies. A detailed investigation will be performed in future work. 

\begin{figure}
    \centering
    \includegraphics[width=\linewidth]{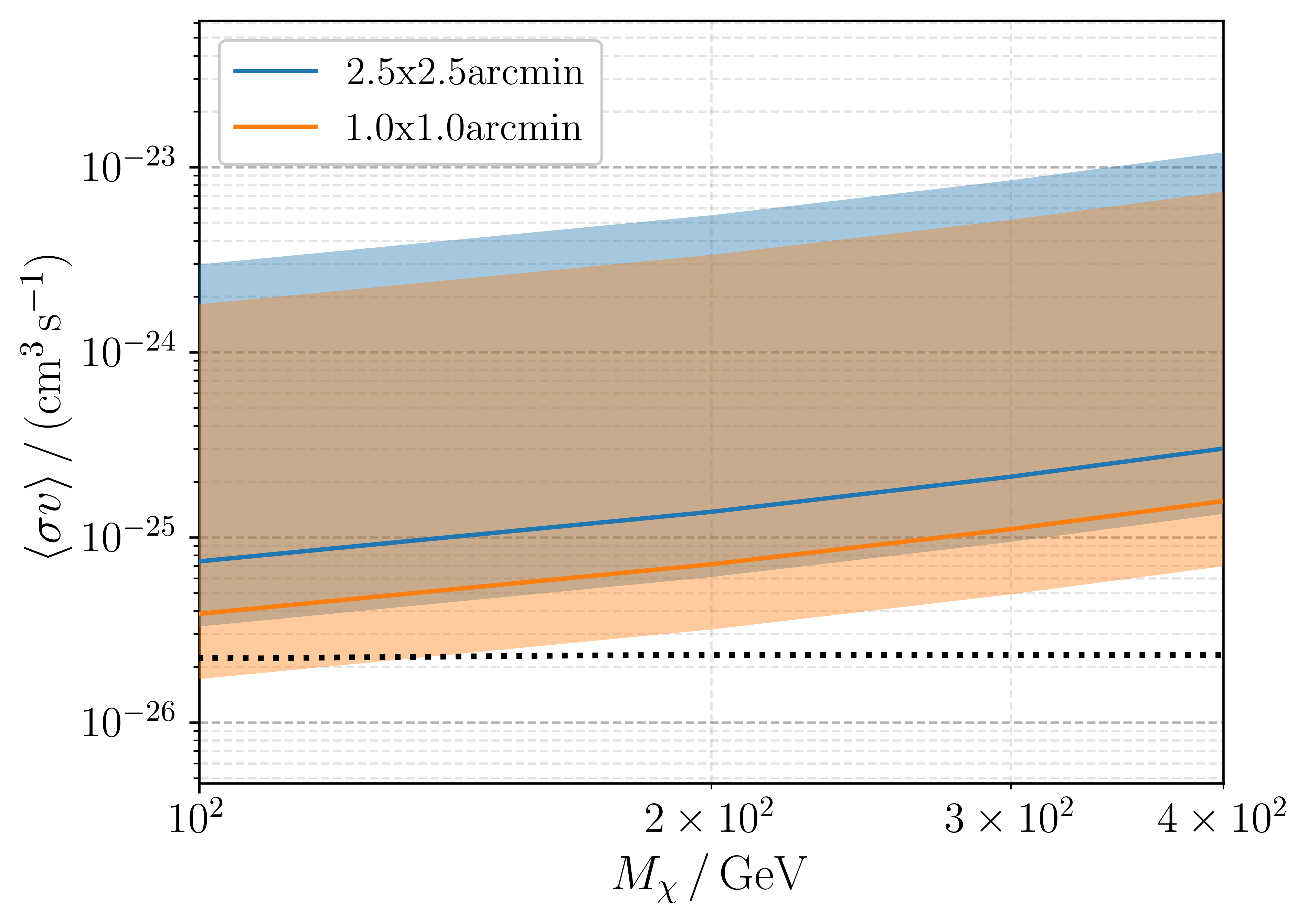}
    \caption{Comparison of the size of chosen RoI. Two sizes are considered, $2.5^{\prime} \times 2.5^{\prime}$ and the reduced region of $1.0^{\prime} \times 1.0^{\prime}$. As an example, we have considered annihilation through bottom quarks in the RXCJ0757.7-5315 cluster. The lower bound of each uncertainty band corresponds to the optimistic case while the upper bound corresponds to the pessimistic case. The thermal relic value is depicted by the dotted black line.}
    \label{fig: size}
\end{figure}

The DM signal is highly concentrated around the modeled central position of the DM density distribution. Thus, it is expected that the central coordinates of the cluster at which the injection is performed will have an effect on the resulting upper limits. As the central position of the galaxy cluster depends on the tracer used there can be a significant offset between the MCXC central positions and the MGCLS central positions. We find that the flux uncertainty method is more robust to changes in the central position, while the pixel-by-pixel method is sensitive to changes. We consider RXCJ0225.1-2928 as an example. The MGCLS central coordinates are (36.3750$^{\circ}$, -29.5$^{\circ}$) and the MCXC central coordinates are (36.293750$^{\circ}$, -29.473889$^{\circ}$). This is an offset of 0.076$^{\circ}$ or 324 kpcs. We find that the cross-section upper limits for the flux uncertainty method vary between these positions by approximately $\sim 1.2$, while for the pixel-by-pixel method, the cross-sections vary by approximately an order of magnitude. This is likely caused by a sensitivity to the surface brightness values of individual pixels within the chosen RoI. Additionally, the size of the RoI has been chosen to ensure that there are a sufficient number of resolution elements for the statistical method to be valid. A $ 2.5^{\prime} \times 2.5^{\prime}$ region has been used in the analysis, however in cases where there are a large number of pixels in the image, a smaller region may be considered. As an example case, when the RoI in the J0757.7-5315 image is reduced to a square of size $1.0^{\prime} \times 1.0^{\prime}$, the annihilation cross-section upper limits are improved by a factor of $\sim 2$, as shown in Figure~\ref{fig: size}.
 
The constraints for the annihilation cross-section for the 2HDM+S DM candidate show significant overlap for the flux uncertainty method. As such, 2HDM+S remains a potentially viable explanation for the various astrophysical excesses investigated by Beck et al \cite{beck2021connecting}. The upper limits derived with the use of MeerKAT allow us to begin to probe the 2HDM+S parameter space. The results presented here are more constraining than those determined with gamma-ray data from Fermi-LAT \cite{albert2017fermi}, as well as the preliminary MeerKAT predictions for Reticulum II determined by Beck et al \cite{beck2021connecting}.

We have presented the annihilation cross-section upper limits for galaxy clusters from MGCLS. Two methods of analysis have been compared, that use the integrated flux and surface brightness values respectively. The integrated flux method appears the more reliable of the two methods, due to the subtraction of the intrinsic noise of the map. The most constraining results are produced for Abell 4038 and RXCJ0225.1-2928 thereafter. In order to determine which properties were crucial for this we investigated annihilation through bottom quarks for Abell 4038. The mass of the dark matter halo was varied through 3 orders of magnitude. Additionally, the gas density parameters were scaled according to Coma properties, in contrast to the values found in the literature. It was found that the most significant change to the upper limits was caused when varying the mass of the dark matter halo. The upper limits were weakened by up to a factor of 15 when the dark matter halo mass was reduced by an order of magnitude. We, therefore, conclude that the more massive halos studied were able to produce more stringent upper limits.  

With the results from Abell 4038, we are able to exclude WIMP values up to approximately $\sim 1000$ GeV for annihilation into bottom quarks. We note that these results assume that the observed synchrotron emission is baryonic in nature. Chan et al \cite{chan2020} fitted the observed synchrotron flux with the various models for the cosmic ray emission. The authors found the largest likelihood values when the data was fit to a \textit{ in situ} acceleration model \cite{chan2020}, which has the form   
\begin{equation}
    S_\mathrm{CR}=S_\mathrm{CR \, 0} \, \left( \frac{\nu}{\mathrm{GHz}}\right)^{-\alpha} \, \mathrm{exp} \left(\frac{-\nu ^{1/2}}{\nu_s^{1/2}}\right) \; ,
\end{equation}
with the free parameters $S_\mathrm{CR \, 0}$, $\alpha$ and $\nu_s$.
The authors found that the total flux was better described when a DM component was added. This resulted in 6$\sigma$ preference for a model containing a DM component to the synchrotron emission for four popular annihilation channels. Our conservative assumption strongly excludes the models in Chan et al \cite{chan2020}.   

In conclusion, we have demonstrated the ability of MeerKAT to constrain the WIMP parameter space by searching for diffuse synchrotron emission in galaxy clusters. These initial results show that MeerKAT has the potential to become the forefront of indirect DM searches. It can be expected that more accurate upper limits will be obtained once detailed polarization studies of galaxy clusters with MeerKAT begin. Our results serve as proof of concept for more sensitive dark matter searches in the upcoming SKA era.

\begin{acknowledgments}
N.L acknowledges the financial assistance of the South African Radio Astronomy Observatory towards this research (www.sarao.ac.za). M.S. acknowledges the financial assistance of the National Research Foundation of South Africa (Bursary No. 112332).

 MeerKAT Galaxy Cluster Legacy Survey (MGCLS) data products were provided by the South African Radio Astronomy Observatory and the MGCLS team and were derived from observations with the MeerKAT radio telescope. The MeerKAT telescope is operated by the South African Radio Astronomy Observatory, which is a facility of the National Research Foundation, an agency of the Department of Science and Innovation. 
\end{acknowledgments}

\bibliography{ref}
\end{document}